\documentclass[aps,twocolumn,raggedbottom,pre,showpacs,nobalancelastpage,amssymb,groupedaddress]{revtex4}
\usepackage{graphicx}
\usepackage{amsmath}

\newcommand{\pll}{\parallel}

\newcommand{\e}{{\rm e}}
\newcommand{\rmd}{{\rm d}}
\newcommand{\rmi}{{\rm i}}

\newcommand{\half}{{\textstyle{\frac{1}{2}}}}

\newcommand{\al}{\alpha}

\newcommand{\de}{\delta}
\newcommand{\De}{\Delta}
\newcommand{\eps}{\epsilon}

\newcommand{\tEo}{\tau_{\rm E}^{\rm op}}
\newcommand{\tEc}{\tau_{\rm E}^{\rm cl}}
\newcommand{\tE}{\tau_{\rm E}}

\newcommand{\tELR}{\tau_{\rm E}^{\rm LR}}

\newcommand{\tD}{\tau_{\rm D}}
\newcommand{\NL}{N_{\rm L}}
\newcommand{\NR}{N_{\rm R}}
\newcommand{\WL}{W_{\rm L}}
\newcommand{\WR}{W_{\rm R}}

\newcommand{\ignore}[1]{\relax}

\begin{document}
\title{Semiclassical Theory of Quantum Chaotic Transport: \\
Phase-Space Splitting, Coherent 
Backscattering and Weak Localization}
\author{Ph.~Jacquod$^{1,2}$ and Robert S. Whitney$^{2,3}$} 
\affiliation{
$^1$ Department of Physics, University of Arizona, 1118 E. 4$^{\rm th}$ Street,
Tucson, AZ 85721, USA
\\
$^2$ D\'epartement de Physique Th\'eorique,
Universit\'e de Gen\`eve, 1211 Gen\`eve 4, Switzerland
\\
$^3$ Institut Laue-Langevin, 6 rue Jules Horowitz, B.P. 156,
      38042 Grenoble, France}
\date{May 10, 2006}
\begin{abstract}
We investigate transport properties of quantized chaotic systems
in the short wavelength limit. We focus on
non-coherent quantities such as the Drude conductance, its
sample-to-sample fluctuations, 
shot-noise and the transmission spectrum, as well
as coherent effects such as weak localization.
We show how these properties are influenced by the emergence 
of the Ehrenfest time scale $\tE$. Expressed in an optimal 
phase-space basis, the scattering matrix acquires a block-diagonal
form as $\tE$ increases, reflecting the splitting of the 
system into two cavities in parallel, a classical deterministic
cavity (with all transmission eigenvalues either 0 or 1)
and a quantum mechanical stochastic cavity. 
This results in the
suppression of the Fano factor for shot-noise and the deviation of
sample-to-sample conductance fluctuations from their universal value.
We further present a semiclassical theory for weak localization which
captures non-ergodic phase-space structures and preserves
the unitarity of the theory. Contrarily to our previous claim 
[Phys.\ Rev.\ Lett.\ {\bf 94}, 116801 (2005)], we find that 
the leading off-diagonal contribution to the conductance leads 
to the exponential suppression of the coherent backscattering peak and of weak
localization at finite $\tE$. This
latter finding is substantiated by numerical magnetoconductance
calculations.
\end{abstract}
\pacs{73.23.-b, 74.40.+k, 05.45.Mt, 05.45.Pq}
\maketitle
\section{Introduction}
Closed chaotic systems are classically characterized by ergodicity,
mixing and a positive Kolmogorov-Sinai (KS) entropy \cite{ll}. 
These three characteristics form a hierarchy: mixing systems are ergodic,
and systems with positive KS entropy are mixing, but the reverse is
not necessarily true.
Ergodicity means that phase-space averages equal time averages,
while the definition of both mixing and KS entropy requires the 
introduction of some phase-space coarse-graining. For mixing,
one needs to define finite-sized phase-space cells inside which
points originating from two initially well separated distributions
of initial conditions are equally likely to be found. As time goes
by, mixing occurs on smaller and smaller scales, i.e.
the minimal volume of these cells decreases. 
The KS entropy
is defined from the measure of the intersection of the cells 
with their back evolution under the system dynamics.
A positive KS entropy means an exponential production of
information, and thus the generation of randomness in the Kolmogorov
sense,
as more and more different trajectories emerge
from apparently indistinguishable initial conditions \cite{ll}. 
For closed systems, the KS entropy
is related to the exponential sensitivity to
initial conditions, and equals the sum of the 
associated positive Lyapunov exponents
\cite{ll,Eckmann}.

The situation becomes different once the system is open and scattering
trajectories are considered. Phase-space structures emerge
then which are prohibited by ergodicity and mixing, even in systems
which have a positive KS entropy when closed. These structures
and their influence on quantum transport
are the focus of this article. We will see how their occurrence
affects transport through open quantized chaotic systems in the
semiclassical, short wavelength limit. They result
in a splitting of the cavity into two cavities in parallel, one 
where transport is ruled by classical determinism
and one where transport exhibits quantum stochasticity.

\subsection{Classical chaos in open systems}

\begin{figure}
\begin{center}
\hspace{-0.5cm}\includegraphics[width=8cm]{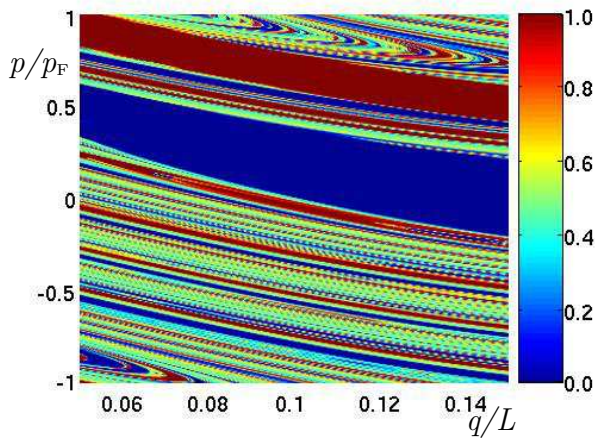}
\end{center}
\vskip -55mm
\hskip -75mm {\large {\it p/p}}$_{\rm F}$
\vskip 43mm
\hskip 45mm  {\large {\it q/L}}
\caption{\label{fig:cl_bands} (Color online) Classical phase-space color
plot of the transmission probability from the phase-space 
projection of the injection 
lead (see Fig.~\ref{fig:sos_def}). The phase-space has been coarse-grained
by a rectangular grid with $9 \cdot 10^4$ cells, 
and the transmission probability
in each cell has been calculated by time-evolving $10^4$ classical trajectories
per cell. Higher transmissions correspond to red, lower
transmissions to blue. The dynamical system used is the open kicked rotator
as defined in Section~\ref{sec:numerics}.}
\end{figure}

\begin{figure}
\begin{center}
\includegraphics[width=8.3cm]{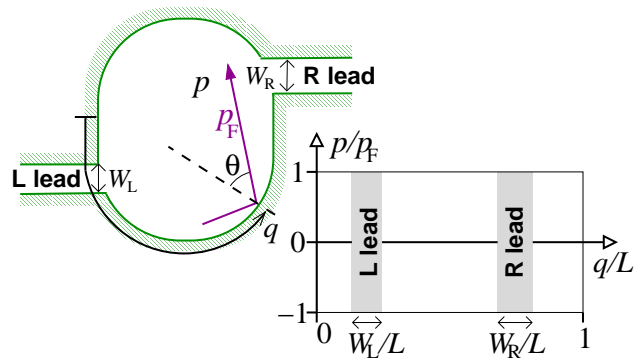}
\end{center}
\caption{\label{fig:sos_def} (Color online) Sketch of a two-terminal 
open chaotic cavity (top left)
and its phase-space represented as 
a Poincar\'e surface of section on the boundary of the cavity (bottom right).
All momenta on the energy surface ($E=p_{\rm F}^2/2 m$) 
are parametrized 
by the tangential momentum $p$ running from $-p_{\rm F}$ to $p_{\rm F}$. 
All possible positions
on the boundary are parametrized by $q$ running from zero to $L$,
where $L$ is the circumference of the cavity. The phase-space
is made dimensionless by normalizing momenta and real-space coordinates
with $p_{\rm F}$ and $L$ respectively.}
\end{figure}

We specialize to two-dimensional chaotic cavities in a two-terminal 
geometry. Typical nonergodic structures occurring in such 
open chaotic systems are illustrated in Fig.~\ref{fig:cl_bands}.
A color plot of the transmission probability is shown on a
phase-space projection of one of the two openings.
The horizontal axis
gives the position on a cross-section of the opening, normalized by the
cavity perimeter $L$, and the vertical axis
gives the momentum component of injection into the system, parallel 
to the cavity boundary, and normalized by the Fermi momentum $p_{\rm F}$.
Both the real-space set-up and the dimensionless phase-space we use
are defined in Fig.~\ref{fig:sos_def}. 

Band-like structures such as those
appearing in Fig.~\ref{fig:cl_bands} have been reported
and discussed earlier
\cite{Wirtz99,Sil03,TwoSN,TwoUCF,Jac04,Goo05}. All scattering trajectories
whose initial point lies in one of the bands have approximately the
same dwell time through the system \cite{caveat1}. 
The typical dwell time $\tau$ of a
band determines its width as $\simeq (W/L) \exp[-\lambda \tau]$ \cite{Sil03}
($W$ is the width of the opening and $\lambda$ is the Lyapunov exponent). 
Thus the largest blue and red bands
in Fig.~\ref{fig:cl_bands} respectively correspond to direct reflection and
transmission, while thinner bands correspond to longer dwell times through
the system. Trajectories inside a band are transported
in one bunch, and the phase-space volume they occupy is blocked for other
trajectories by Liouville's theorem. Because trajectories
remain inside the system for a finite time, the definition of
ergodicity, that 
\begin{equation}\label{eq:ergodicity}
\Omega^{-1} \int_{\Omega} {\rm d} {\bf p} {\rm d} {\bf q} \; f({\bf p},{\bf q};t) = 
\lim_{\tau \rightarrow \infty} \tau^{-1} \int_0^\tau 
{\rm d} t f({\bf p}_0,{\bf q}_0;t),
\end{equation}
for almost all functions $f({\bf p},{\bf q};t)$ and almost all phase-space
points $({\bf p}_0,{\bf q}_0) \in \Omega$ 
no longer holds, but depends on
$({\bf p}_0,{\bf q}_0)$. The time-integral on the right-hand side
of Eq.~(\ref{eq:ergodicity}) extends only up to the dwell time $t_0$ of
the one trajectory going through $({\bf p}_0,{\bf q}_0)$, and accordingly
Eq.~(\ref{eq:ergodicity}) cannot be
preserved over the full phase-space $\Omega$.
Simultaneously, mixing occurs on a given
scale  only for the subset of trajectories longer than 
some finite dwell time. Scattering trajectories through open systems have 
a continuous distribution of dwell times $P(t)$
and because of the exponentially decreasing volume of 
scattering bands, mixing occurs on exponentially
smaller scales on longer trajectories. Reversing the argument,
a given phase-space resolution volume $\xi$ corresponds to a time scale
$\tau_\xi \approx \lambda^{-1} \ln[(W/L)^2/\xi]$. 
Long trajectories with $\tau>\tau_\xi$ exhibit mixing on the scale $\xi$,
while short trajectories with $\tau<\tau_\xi$ lie on bands well resolved
by $\xi$-cells, which therefore do not have the mixing property.
These two sets of classical scattering trajectories have no
phase-space overlap.

\subsection{Quantum chaos in open systems}

\begin{figure*}
\begin{center} 
\hspace{-0.7cm}\includegraphics[width=6.25cm]{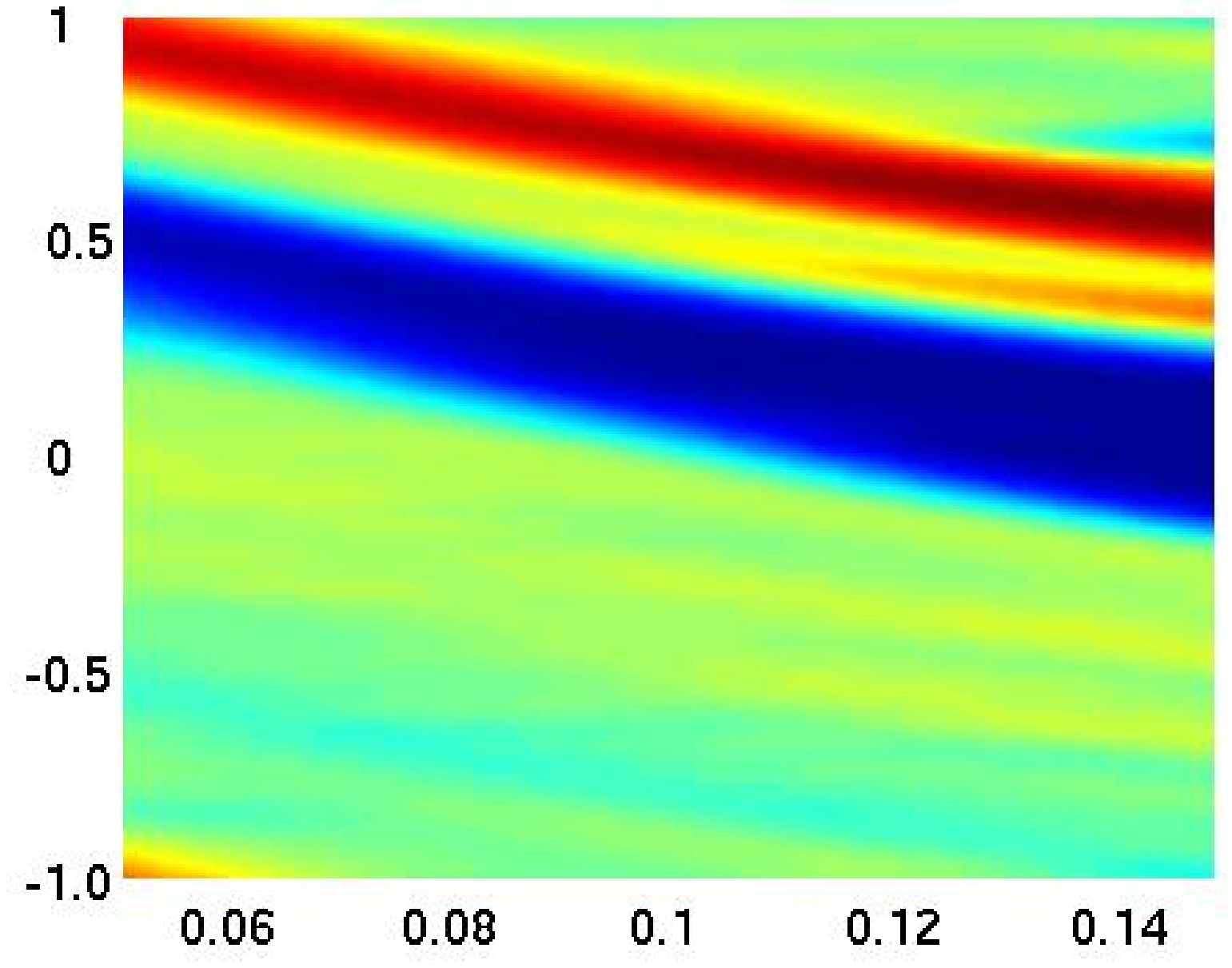}
\hspace{-0.45cm}\includegraphics[width=6.25cm]{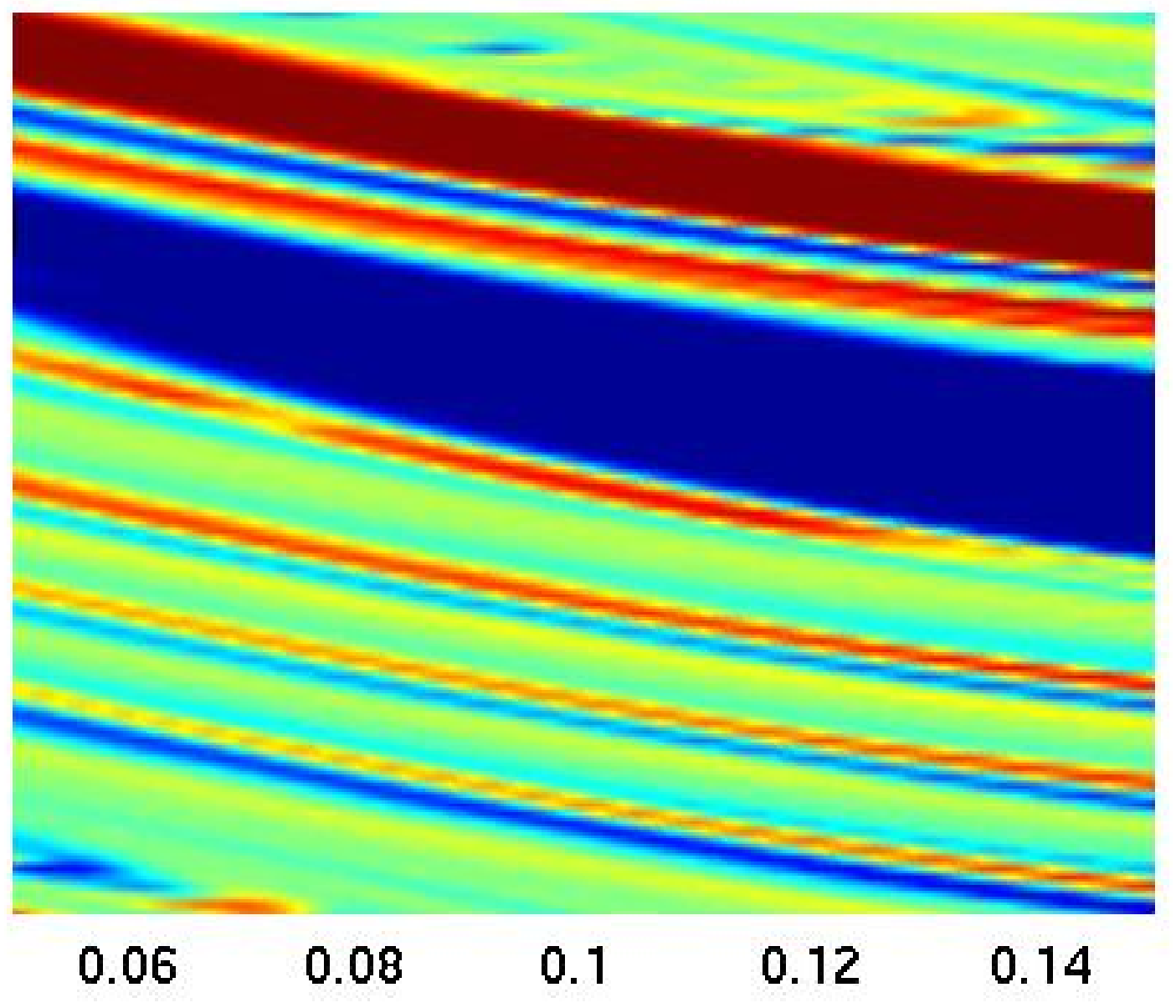}
\hspace{-0cm}\includegraphics[width=6.25cm]{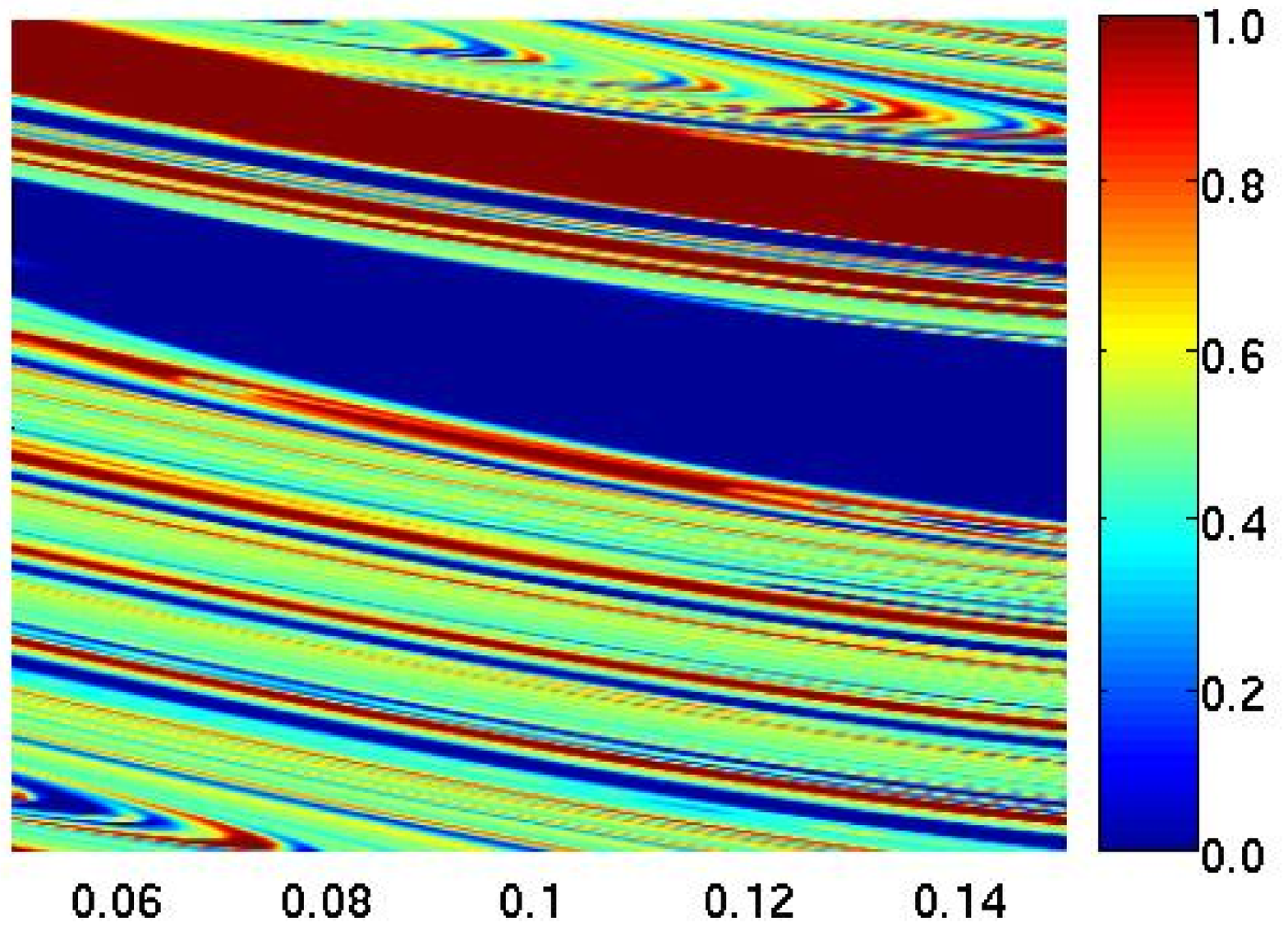}
\end{center} 
\caption{\label{fig:qm_bands} (Color online) Quantum phase-space color
plots of the transmission probability from the injection lead.
The system is the quantum equivalent of the classical system of 
Fig.~\ref{fig:cl_bands}. The phase-space has been coarse-grained
by a rectangular grid with $9 \cdot 10^4$ points. Starting from  
each point of the grid, an isotropic Gaussian wave-packet has been time-evolved
and its transmission probability calculated. From left to right, the three
panels correspond 
to decreasing effective Planck's constant $\hbar_{\rm eff} = 2 \pi/M$ with
$M=512$ (with a conductance $g=22.4$ and a Fano factor $F=0.193$), 
8192 ($g=375.9$ and $F=0.121$) and 131072 ($g=5990.8$ and $F \approx 0.08$)
respectively. 
More and more fine-structured details of the classical phase-space
are resolved as $\hbar_{\rm eff} \rightarrow 0$.
Higher transmissions correspond to red, lower
transmissions to blue color.}
\end{figure*}

A finite resolution scale emerges naturally when the system is quantized:
the phase-space becomes tiled with cells of volume $2 \pi \hbar$.
For particles with a finite Fermi wavelength  
$\lambda_{\rm F}$, this is equivalent to tiling the dimensionless phase-space
projection of the leads (see Fig.~\ref{fig:sos_def}) with cells of volume 
$\hbar_{\rm eff}=\lambda_{\rm F}/L$, the effective Planck's constant.
This leads to the existence of a finite number 
$N={\rm Int}[2 W/\lambda_{\rm F}]$
of conduction channels through the system. As $\hbar_{\rm eff}$ is
made smaller and smaller, all classical parameters being kept constant
(the semiclassical limit), more and more of the band structures of 
Fig.~\ref{fig:cl_bands} are resolved (see Fig.~\ref{fig:qm_bands}). 
Conversely, more and more
of the conduction channels are supported by one and only one
classical transmission or reflection band, and thereby become deterministic.
It is thus natural to investigate the effect that the lack of
mixing of short trajectories has on properties
of open quantum chaotic systems.

Transport through ballistic quantum cavities, so-called quantum dots,
has been investigated intensively in recent years \cite{Revdot}. 
In the regime where the dot's size is much larger than the 
Fermi wavelength, $L \gg \lambda_{\rm F}$, transport has been shown to
depend on the integrability or lack thereof of the 
classical dynamics, as determined 
by the confinement potential \cite{Bar93,Mar93}.
Most experimental investigations so far
have focused on the limit of few, $\lesssim 10$ conduction channels,
where it has been found that
quantum transport in the chaotic case exhibits a
universality which is well captured by the Random Matrix Theory (RMT)
of transport \cite{Meh91,RMT}. Recently, a semiclassical
theory for the conductance of ballistic cavities has been
developed \cite{haake}, confirming the common belief that 
RMT universality applies at least to a certain regime of ballistic chaos.

It is however well-known that physically relevant time scales
restrict the range of validity of RMT.
In recent years it has become clear that the {\it Ehrenfest} 
time $\tE$ does this in ballistic quantum chaotic systems, 
with RMT ceasing to be valid when $\tE$ becomes relevant.
The Ehrenfest time is the time it takes
for the chaotic classical dynamics to stretch a narrow
wave packet, of spatial extension $\lambda_{\rm F}$, to some
relevant classical length scale ${\cal L}$. Since the stretching is
exponential in a chaotic system, 
one has $\tE=\lambda^{-1}
\ln[{\cal L}/\hbar_{\rm eff} L]$ \cite{Berman78}.
The scattering of an initially localized wavepacket into
all possible modes (similar to $s$-wave scattering on a restricted
portion of phase-space) is only established after
classical mixing has set in on the scale $\hbar_{\rm eff}$, i.e.
for times longer than $\tE$. For
shorter times, the quantum dynamics is deterministic. One thus expects
deviations from RMT to emerge as $\tE/\tD$ increases.

This line of reasoning has been qualitatively confirmed in the cavity
transport experiments of Ref.~\cite{Obe02}, which
observed a significant
reduction of the electronic shot-noise power below its
RMT value upon opening the cavity more and more.
This reduction is very likely 
due to an increasing fraction of deterministic channels
in the transmission spectrum of the cavity which can be understood as follows.
The shot-noise power (the intrinsically quantum part of the fluctuations
of the electronic current) 
is usually quantified by the dimensionless Fano factor $F$, 
the ratio of the shot-noise to the Poissonian noise \cite{Bla00},
which can 
be expressed in term of the transmission
spectrum $T_i \in [0,1]$ of the cavity as \cite{Bla00}
\begin{equation}\label{eq:fano}
F =\sum_i T_i (1-T_i) \Big/ \sum_i T_i.
\end{equation}
Hence deterministic channels, those having $T_i=0$ or 1, do not
contribute to $F$. Such channels appear as the classical 
bands discussed above are quantum mechanically resolved, which
can be achieved either by reducing the dwell time or by reducing
the Fermi wavelength. The former change was more appropriate for the
experimental set-up of Ref.~\cite{Obe02}, and the latter change
is illustrated in the numerics of
Fig.~\ref{fig:qm_bands}. We show three quantum phase-space plots for a
fixed classical set-up (the same as for Fig.~\ref{fig:cl_bands}).
Plotted is the quantum transmission probability 
$\langle (x,p) | {\bf T} | (x,p) \rangle$ for a fixed
grid of initial coherent states $| (x,p) \rangle$, 
i.e. isotropic Gaussian wavepackets centered 
on $(x,p)$. The three panels from left to
right correspond to smaller and smaller $\lambda_{\rm F}$. It is seen that
as $\lambda_{\rm F}$ decreases, 
finer and finer phase-space structures are resolved. Moreover, 
coherent states entirely lying on deep red (deep blue) regions have
$\langle (x,p) | {\bf T} | (x,p) \rangle=1$ (0), and are
therefore eigenstates of ${\bf T}$. Any of them can carry a quantum
channel which does not contribute to shot-noise
(the total number of deterministic channels is obtained 
only after the orthogonalization of the coherent states, see below).
With decreasing $\lambda_{\rm F}$, the number 
of deterministically transmitted coherent states 
increases faster than the total number of channels, 
inducing the reduction of the shot-noise power below its RMT value.

\begin{figure}
\includegraphics[width=6cm]{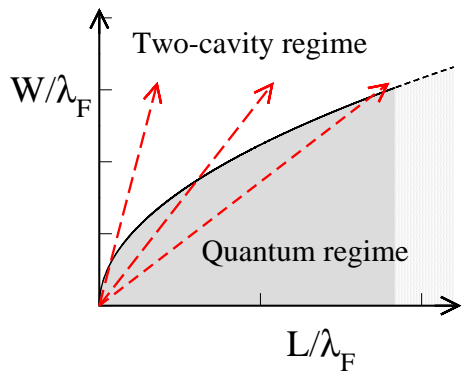}
\caption{\label{fig:regimes} (Color online)
Schematic of the different transport regimes through a ballistic 
chaotic cavity with perimeter $L$ coupled to leads of width $W$.
Above the separatrix, the system splits into two effective cavities,
one purely classical and the other quantum mechanical. Only the latter
contains quantum effects such as non-deterministic transmission and
quantum interferences. 
Below the separatrix $W=(\lambda_{\rm F}L)^{1/2}$ (solid curve), 
all modes are quantum mechanical (grey area).
The red dashed arrows indicate the semiclassical
limit of $\hbar_{\rm eff} \rightarrow 0$ at fixed classical parameters.
The slope of the arrows is given by the inverse dwell time $\tD$.}
\end{figure}

The suppression of the Fano factor in the semiclassical limit 
was anticipated long ago \cite{vanhouten}.
More recent works quantitatively 
predicted a suppression $F \propto \exp(-\tau_{\rm
  E}/\tau_{\rm D})$, in term of the new Ehrenfest time scale
and the average dwell time $\tau_{\rm D}$ through the system
\cite{Aga00,wj-fano}, a suppression which was related to
the phase-space resolution picture of Ref.~\cite{Sil03} given above
and confirmed numerically in Refs.~\cite{TwoSN,Jac04}. Ref.~\cite{Whi05}
presented a phase-space semiclassical approach resolving the classical bands
which showed that the fraction of deterministic transmission eigenvalues
not contributing to noise 
is $\propto [1-\exp(-\tE/\tau_{\rm D})]$.
Following the numerous recent investigations of the
quantum-to-classical correspondence in open systems, which we now proceed
to briefly summarize, it has become clear that 
$\tE/\tD \rightarrow 0$ is a necessary condition for
complete RMT universality \cite{Henning05}. As is illustrated 
in Fig.~\ref{fig:regimes}, this condition is never satisfied in the 
semiclassical limit $\hbar_{\rm eff} \rightarrow 0$.

Following Ref.~\cite{Ale96} which suggested that the existence 
of a finite  $\tE$ discriminates quantum chaotic from quantum disordered
systems (with the former class exhibiting
deviations from universality), many investigations have been 
devoted to the study of open quantum chaotic systems at
finite $\tE$ \cite{Henning05}.
Focusing on transport, it is by now well established numerically that, 
as $\tE/\tau_{\rm D} \rightarrow \infty$, 
the Fano factor disappears \cite{TwoSN}, and
sample-to-sample
conductance fluctuations lose their universality.
Simultaneously,
parametric conductance fluctuations appear to remain universal 
\cite{Jac04,TwoUCF}.
The weak localization correction to the conductance was first predicted
to disappear exponentially as $\delta g \propto \exp[-2 \tE/\tD]$ \cite{Ale96}
or $\delta g \propto \exp[-\tE/\tD]$ \cite{Ada03}. Early
numerical works, on the other hand, concluded that 
$\delta g$ is independent of $\tE/\tD$ \cite{TwoWL}.
This puzzle was resolved in Ref.~\cite{Bro05} (though some skepticism
remained, on the part of the authors of the present article amongst others)
whose analytical approach predicted $\delta g \propto \exp[-\tE/\tD]$
in agreement with numerical simulations. Below we will present both an
analytical, semiclassical treatment of weak localization with
a special emphasis on backscattering and unitarity,
and numerical magnetoconductance data giving a microscopic confirmation
of the conclusion $\delta g \propto \exp[-\tE/\tD]$ of Ref.~\cite{Bro05}.

There are currently several theories for open quantum chaotic systems at
finite $\tE$. First, the stochastic quasiclassical theory 
mimics the post-Ehrenfest time mode-mixing by introducing
fictitious random scatterers with a scattering rate appropriately tuned
\cite{Aga00,Ale96,Bro05}. It is
developed from standard methods in disordered systems, but
breaks time-reversal symmetry at the classical level
already. 
Second, there is a semiclassical theory \cite{Ada03},
     which until now had not been shown to preserve the unitarity 
     of the scattering matrix (and hence conserve the current).
Third, a phenomenological model 
originating from Ref.~\cite{Sil03} models the total electronic fluid
as two coexisting phases, a classical and a quantum one. At this level, 
the theory is referred to as the two-phase fluid model \cite{Jac04,Whi05}. 
With the additional surmise made in Ref.~\cite{Sil03} that the quantum
phase has RMT properties, one gets the effective RMT model. 
The effective RMT model successfully explains the behavior of
shot-noise, the transmission spectrum and conductance fluctuations,
but is in contradiction with the 
disappearance of weak localization \cite{Bro05}. 
The suppression of weak localization at large
$\tE/\tD$ invalidates the effective RMT model, but not
the two-phase fluid model, as we will explicitly see below. 

\subsection{Outline of this article}

In this article, we focus on
quantities such as the average conductance, 
shot-noise and the transmission spectrum in ballistic chaotic
cavities, as well as the weak localization correction to the 
conductance as $\hbar_{\rm eff} \rightarrow 0$. 
These quantities
are strongly influenced by the emergence of 
the open cavity Ehrenfest time
$\tEo = \lambda^{-1} \ln[(W/L) ^2/2 \pi \hbar_{\rm eff}]$ \cite{Vav03}. 
All classical parameters
being fixed, that limit inevitably turns any system into a nonuniversal
quantum chaotic one as $\tEo$ becomes finite (see Fig.~\ref{fig:regimes}). 
We calculate the scattering matrix
in a basis that optimally resolves phase-space structures and show that the
system splits into two cavities in parallel. This provides 
a foundation for the two-phase fluid model. 
We go significantly beyond our previous work, Ref.~\cite{Whi05}, with 
(i) a detailed construction of a basis
which optimally resolves those phase-space structures and (ii) an
explicit semiclassical calculation of the weak localization correction
to the conductance which preserves the unitarity of the scattering matrix 
to leading order. 

The outline of the paper goes as follows. In Section~\ref{sect:classical},
we discuss the nonergodic classical structures of open chaotic systems
such as those shown in Fig.~\ref{fig:cl_bands}
and the Ehrenfest time scale that accompanies them.
Our task requires that we resolve quantum mechanically these classical
phase-space structures. This suggests that we
employ a semiclassical theory. The existing such theories
\cite{haake,Bar93} have to be replaced by a phase-space
resolving theory, which requires the construction of 
an appropriate orthogonal mode
basis. This is done is Section~\ref{sect:ps-basis}. We then
write the system's scattering matrix in this basis, and show how
this results in phase-space splitting at the quantum level in Section 
~\ref{sect:S-and-T-in-ps-basis}. 
Amplifying on that, we show how
deterministic modes emerge 
and calculate the average Drude conductance and its sample-to-sample
fluctuations at large $\tEo$. 
In Section~\ref{sect:weak-loc}, we present our semiclassical theory for
the weak localization correction to the conductance. Several key aspects 
absent in previous semiclassical treatments are stressed
here. In particular we present the first semiclassical calculation of the 
coherent backscattering peak
at finite $\tEo/\tD$. We show that both coherent backscattering and
weak localization are exponentially suppressed $\propto \exp[-\tEc/\tD]$
with the closed cavity Ehrenfest time 
$\tEc=\lambda^{-1} \ln[\hbar_{\rm eff}^{-1}]$.
This is so, because weak localization and coherent 
backscattering come from trajectories longer than $\tEo+\tEc$, which have an
exponentially small relative weight $\exp[-\tEc/\tD]$ in the stochastic
block of the scattering matrix. 
The existence of two separated fluids is however confirmed. 
We demonstrate that the classical phase-space structures
(which give rise to
phase-space splitting) must be included in the semiclassical treatment
to preserve the unitarity of the scattering matrix 
at nonvanishing value of $\tEo/\tD$. 
We also point out the origin of the discrepancy between 
our final conclusion, that $\delta g \propto \exp[-\tEc/\tD]$, and our earlier
claim that $\delta g$ remains at its universal value in the
deep semiclassical limit \cite{Whi05}. 
We finally present numerical magnetoconductance data confirming our theory. 
Summary and 
conclusions are presented in Section~\ref{sect:conclusions}, 
and technical details discussed in the Appendices.

\section{Classical Scattering Structures and Ehrenfest times}
\label{sect:classical}

\subsection{Transmission and reflection bands}
\label{sect:trans-and-refl-bands}

We consider classical scattering trajectories. They are injected 
into the cavity from one of the two leads, say the left (L) lead, 
with initial position $q$ and momentum $p$ on a cross-section of the lead.
The momentum is on the Fermi energy surface $E=p_{\rm F}^2/2m$. 
The trajectory is determined by ballistic motion inside the confinement
potential defining the cavity, until the particle hits
the boundary between the cavity and one of the leads, at what time it
escapes. Throughout this paper we will consider perfectly
transparent leads.

\begin{figure}
\includegraphics[width=8cm]{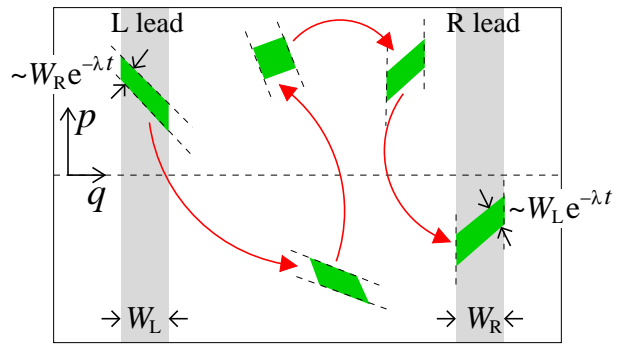}
\caption{\label{fig:band} (Color online)
Time evolution of an L to R transmission band (parallelograms) 
on the boundary of the cavity of Fig.~\ref{fig:sos_def}. 
The leads are indicated by the shaded rectangles.}
\end{figure}

Scattering trajectories are not isolated,
instead they occur in bands (see Fig.~\ref{fig:cl_bands}).
As mentioned in the introduction, a scattering band is a phase-space structure 
occurring in open systems, even when their closed counterpart is fully
chaotic. It contains a set of trajectories which exit at approximately 
the same time through the same lead \cite{caveat1}, 
having followed similar  trajectories through the cavity,
in the sense that any trajectory in the band can be topologically deformed 
into any other. The situation is sketched in Fig.~\ref{fig:band}.
Bands on the injection lead are defined by the overlap of that
lead with the time-reversed evolution of the exit lead, including 
absorption at both leads.

For an individual system, the exact number and area of the bands 
depends on the specifics of the lead positions and widths, and the
cavity shape. However, averaged properties calculated over an ensemble 
of cavities with the same dwell time and Lyapunov exponent can be calculated. 
The asymptotic average survival probability is exponential
\cite{Bauer90},
\begin{eqnarray}\label{eq:rhotd}
\rho(\tau) = \exp[-\tau/\tau_{\rm D}].
\end{eqnarray} 
It depends solely on the average dwell time $\tD$
given by 
\begin{eqnarray}
\tau_{\rm D} = \frac{\pi A}{v_F ( W_{\rm L} + W_{\rm R})}.
\end{eqnarray}
Here $A$ is the area of the cavity and
we considered leads with different widths $W_{L,R}$.

In the dimensionless phase-space defined in Fig.~\ref{fig:sos_def},
where momenta and distances are measured in units of $p_{\rm F}$ 
and $L$ respectively,
the injection lead (we always assume this is the L lead)
has a dimensionless phase-space area of
\begin{eqnarray}
\Sigma_{\rm L} = 2 W_{\rm L}/L.
\end{eqnarray}
The fraction of the lead phase-space which couples
to transmitting trajectories is $\sim W_{\rm R}/(W_{\rm L}+W_{\rm R})$,
with the rest coupling to reflecting trajectories.
The average phase-space area
of a single transmission (L$\to$R) 
or reflection (L$\to$L) band which exits at time $\tau$,
is given by 
\begin{eqnarray}
\langle \Sigma_{{\rm L}\to K} (\tau) \rangle
&\sim& {W_{\rm L} W_K \over L^2} \e^{-\lambda \tau},
\label{eq:band-area}
\end{eqnarray}
where $K = {\rm L}, {\rm R}$.
The average number $\langle n_{{\rm L}\to K}(\tau)\rangle$ 
of bands exiting at time $\tau$ through lead $K$ is given by
multiplying
$\Sigma_{\rm L}/ \langle\Sigma_{{\rm L}\to K}(\tau)\rangle$ 
by the probability $(W_K/L) \exp [-\tau/\tau_{\rm D}]$
to escape through the $K$ lead at $\tau$. Hence one has
\begin{eqnarray}
\langle n_{\rm L\to L} (\tau)\rangle= \langle n_{\rm L\to R} (\tau) \rangle
&\sim&
\exp [\lambda \tau (1 - (\lambda \tau_{\rm D})^{-1})].
\qquad
\end{eqnarray}
Since we assume fully developed chaos, $\lambda\tau_{\rm D} \gg 1 $,
we see that the average number of bands diverges as $\tau$ goes to infinity
even though the sum of their phase-space areas goes to zero.
We also note that the average 
number of reflection and transmission bands are equal, 
with only their areas being dependent on $W_{\rm L,R}$.

\subsection{Ehrenfest times and modes on classical bands}
\label{sect:Ehren-times}
The Ehrenfest time scale emerges out of the quantum mechanical 
coarse-graining of phase-space and the partial resolution
of scattering bands. It is the time it takes for quantum mechanical
uncertainties to blow up to some relevant classical scale ${\cal L}$ 
in chaotic systems. The scale ${\cal L}$ depends on the problem at hand,
e.g. on whether the system is closed \cite{Berman78},
or open \cite{Vav03,Scho04}. For the transport set-up we
will focus on, this scale is related to the area of scattering bands.
Large scattering bands, 
those with phase-space area greater than $2\pi \hbar$, 
can carry a number of modes of order their phase-space area 
divided by $2\pi \hbar$. All those modes are classical,
deterministic and exhibit no quantum effects. They are supported by
trajectories shorter than the Ehrenfest time.
The small bands on the other hand, 
those with area less than $2\pi \hbar$ carry less than a full mode,
which generates quantum (stochastic) modes, sitting on many small bands with
dwell times longer than the Ehrenfest time, and hence being partially
transmitted and partially reflected.
Eq.~(\ref{eq:band-area}) then defines 
two open cavity Ehrenfest times for states injected from the L lead, 
one for transmitting bands and one for reflecting bands
\begin{eqnarray}
\tau_{\rm E}^{{\rm L}K}
\,=\, \lambda^{-1} 
\ln \left[ {W_{\rm L}W_K \over 2\pi\hbar_{\rm eff} L^2 }\right]
\;\;\; ; \;\;\; K={\rm L,R}.
\label{eq:tau_E^LK}
\end{eqnarray}
The difference between $\tE^{\rm LR}$ and 
$\tE^{\rm LL}$ is only logarithmic in $W_{\rm R}/W_{\rm L}$.
We will often neglect it and consider instead the 
symmetric open cavity Ehrenfest time 
$\tE = \lambda^{-1} \ln[(W/L)^2/2\pi\hbar_{\rm eff}]$.
The open cavity Ehrenfest times, $\tE^{{\rm L}K}$ 
can be interpreted as the time
it takes for a wavepacket of width $W_{\rm L}/L$ along the stable
manifold of the hyperbolic classical dynamics to evolve into 
a wavepacket with width $W_K/L$ in the unstable direction.

We can readily estimate the number of quantum scattering modes.
The proportion of the L lead phase-space which couples to trajectories
to the $K$ lead with $\tau > \tE^{{\rm L}K}$ is on average
$\e^{-\tE^{{\rm L}K}/\tD} W_K/(W_{\rm L}+W_{\rm R})$.
Thus the average number of quantum modes in the L lead is
\begin{eqnarray}
\langle N^{\rm qm}_{\rm L} \rangle
&=&
N_{\rm L}{N_{\rm L}\e^{-\tE^{\rm LL}/\tau_{\rm D}}
+ N_{\rm R}\e^{-\tE^{\rm LR}/\tau_{\rm D}} 
\over N_{\rm L} +N_{\rm R}} .
\label{eq:Nqm_L}
\end{eqnarray}
All other modes of the L lead are in transmission bands with 
$\tau < \tE^{\rm LR}$ or reflection bands with 
$\tau < \tE^{\rm LL}$,
and so they are all classical modes.
Their number is thus 
\begin{eqnarray}
\langle N^{\rm cl}_{\rm L}\rangle
&=& N_{\rm L} - \langle N^{\rm qm}_{\rm L} \rangle 
\nonumber\\
&=& N_{\rm L}(1-\e^{-\tE^{\rm LR}/\tau_{\rm D}})
+{\cal O}[(\lambda \tD)^{-1}].
\label{eq:Ncl_L}
\end{eqnarray}
The counting argument leading to these estimates finds a 
rigorous derivation below
in section \ref{sect:optimal-ps-basis}, where we 
explicitly cover scattering bands with an orthonormal phase-space
(PS) basis. There, we
also identify a third class of states, overlapping significantly
but still only partially with large bands 
with $\tau < \tE^{{\rm L}K}$. Because of their small relative number
however, these states only have 
a subdominant effect on the system's properties.

Note that the number of classical PS-states 
goes like $\hbar_{\rm eff}^{-1} (1-\hbar_{\rm eff}^{-1/(\lambda \tD)})$, while
the number of quantum PS-states goes like 
$\hbar_{\rm eff}^{-1-1/(\lambda \tD)}$.
In the semiclassical limit $\hbar_{\rm eff}\to 0$
we see that the number of quantum PS-states goes to infinity,
while their fraction goes to zero.

\section{The Phase-Space Basis}
\label{sect:ps-basis}

In order to formally split classical from quantum modes,
our task now is to construct a complete orthonormal basis resolving maximally
the scattering band structure of the classical phase-space. This requires
an optimal resolution in both real-space and momentum coordinates. To 
achieve that, we use results from wavelet analysis. 

\subsection{Existence of orthogonal phase-space bases}
\label{sect:wavelet}

\begin{figure}
\includegraphics[width=8cm]{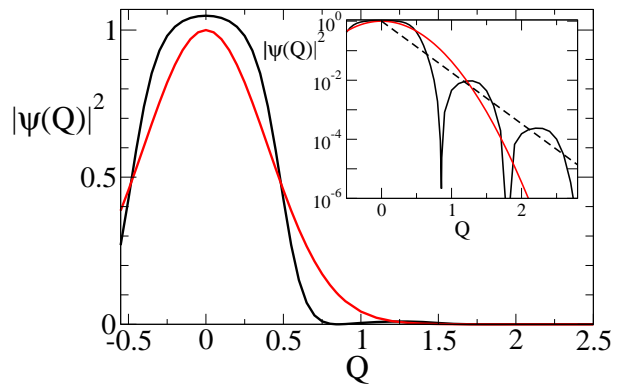}
\caption{\label{fig:ps-state} (Color online)
Main plot: Plot of the real-space wavefunction components 
$|\langle Q |{\rm ps};0,0\rangle|^2$ of a PS-basis state (black line),
and $|\langle Q |{\rm cs};0,0\rangle|^2$ of a coherent state (red line).
Both wavefunctions are symmetric in the dimensionless phase-space
under $Q \leftrightarrow P$, up to a scaling factor.  
Inset: Logarithmic plot of $|\langle Q |{\rm ps};0,0\rangle|^2$,
showing that the PS-state decays exponentially in position (dashed
line). The oscillations in $|\langle Q |{\rm ps};0,0\rangle|^2$ 
ensures it is orthogonal to PS-states centered at finite $Q$,
while its broader central peak
ensures that it is orthogonal to PS-states centered at finite $P$.}
\end{figure}

The existence of complete orthonormal
bases with states exponentially localized in
time and frequency has been proven in the context of
wavelet analysis \cite{wavelet-books}.  
We use such a basis as a PS-basis,
in which each basis state is exponentially localized in position 
and momentum.
We are unaware of any such basis which has closed form 
expressions for the basis states. There are however numerous
algorithms for generating such bases \cite{wavelet-books}.
In Appendix \ref{appendix:example-ps-basis} we
give such an algorithm which
iteratively orthogonalizes a complete but non-orthogonal
basis of coherent states,
generating a set of PS-states of the form shown in 
Fig.~\ref{fig:ps-state}.
While we give this explicit example, 
we emphasize that our theory only requires the existence of such a basis.
We use the fact that each basis state is exponentially localized
in position and momentum,
and that any such complete orthonormal basis
remains complete and orthonormal under any
rotation, translation or area-preserving stretch in phase-space.
Having constructed the PS-basis, the transformation
which takes us from lead modes
      to PS-states is unitary since both bases are orthonormal.

\subsection{The optimal phase-space basis}
\label{sect:optimal-ps-basis}

\begin{figure}
\includegraphics[width=8cm]{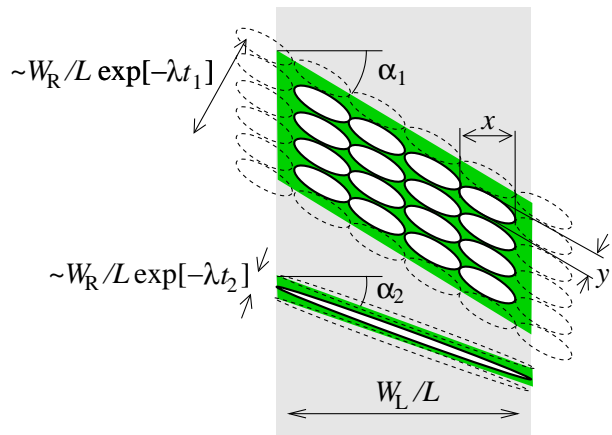}
\caption{\label{fig:band-and-ps-states} (Color online)
Sketch of two parallelogram bands (green areas)
with the PS-states superimposed on them (ellipses).
The upper band has a phase-space area which is a bit more than sixteen times
$2\pi \hbar$, while the lower band has a phase-space area a bit more than
$2\pi \hbar$.
In both cases the basis is optimized, the 
lattice of PS-states is stretched and rotated 
such that the maximum number of PS-states
can be fitted into each band (solid-edged ellipses) 
with the minimum number partially in the band (dashed-edged ellipses).  
Note that these optimally chosen
PS-states have the same aspect ratio as the classical band in which they sit,
thus their longitudinal and transversal extensions,
$x$ and $y$, are given by Eq.~(\ref{eq:aspect-ratio}).}
\end{figure}

\begin{figure}
\includegraphics[width=3.5cm]{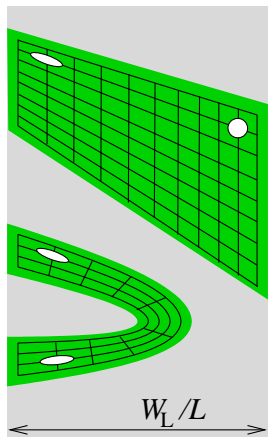}
\caption{\label{fig:trap-band} (Color online)
Sketch of a trapezoidal (top green area) and a folded (bottom
green area) scattering band with a lattice of PS-states superimposed on them.
Both bands cover a phase-space area larger than $2\pi \hbar$.
In both cases the basis is optimized, the 
lattice of PS-states is locally stretched and rotated 
such that the maximum number of PS-states
can be fitted into each band, some of them are indicated by 
solid-edged ellipses.}
\end{figure}

In a recent letter \cite{Whi05}, 
we constructed an orthonormal PS-basis on a square von Neumann lattice.
This basis is simple to explain and work with,
however, it underestimates the number of classical modes
and in particular leads one to predict that the open cavity Ehrenfest time 
is half its correct value \cite{footnote:square-ps-basis}.
To obtain the correct value of $\tE$, the von Neumann lattice
must be adapted to fit in the classical band structures as best as it can.
This is done band by band. For parallelogram bands, the procedure is
to rotate and stretch the originally square von Neumann lattice to a 
parallelogram von Neumann lattice. This is illustrated in
Fig.~\ref{fig:band-and-ps-states}. Each lattice cell still covers an area
$2 \pi \hbar$ and the intraband orthogonality is ensured.
The interband basis orthogonality is preserved due to 
the exponentially small overlap of PS-basis states in different bands
(classical bands do not overlap thanks to Liouville's theorem; 
this effect has been termed {\it Liouville blocking} in Ref.~\cite{Whi05}),
except for a minority of states lying directly at the boundary of the band 
which we will deal with below.
This procedure can still be applied as long as the band's curvature 
is not too large, or for bands which look more trapezoidal than 
parallelogram-like. All one needs to do then is adapt locally the aspect ratio
of the von Neumann lattice, as shown on Fig.~\ref{fig:trap-band}.
Bands with small curvatures dominate at short dwell times. 
However, some bands with larger dwell times 
inevitably display a fold. For those bands, the procedure is to bend the
von Neumann lattice along the axes defined by the eigenvectors
of the stability matrix of the classical dynamics at each point in the band's
phase-space, as in Fig.~\ref{fig:trap-band}. 
The aspect ratio of the lattice is chosen so it obeys
Eq.~(\ref{eq:aspect-ratio}) locally. 
For
intermediate values of $\hbar_{\rm eff}$, the local curvature of the resulting
lattice destroys the orthogonality of the PS-states, however, as
$\hbar_{\rm eff}$ is reduced further and further, the curvature drops out of
the problem, any smooth curve being locally well approximated by a straight 
line. To formally show that an optimal orthogonal basis can be generated
from the square von Neumann basis of Ref.~\cite{Whi05} it is thus
sufficient to (i) consider parallelogram and trapezoidal bands only, 
keeping in mind however
that (ii) at any finite value
of $\hbar_{\rm eff}$, deviations from parallelogram shape generates 
additional non-optimal PS-basis states. The latter result from a further
orthogonalization (e.g. Gram-Schmidt)
required for edge-of-band states and for states on folds
whose curvature is 
not yet well resolved at this value of $\hbar_{\rm eff}$. Below in
Appendix~\ref{sect:appendix-eob-states} we will see that those states 
build up a negligible fraction $N_{\rm qm}/(\lambda \tD) \ll
N_{\rm qm} \ll N_{\rm cl}$ of the total number of modes
in the semiclassical limit. The completeness of the basis
follows from the orthogonality and the conservation of the total number
of basis states, the above procedure being area-conserving.
We are now ready to extend the discussion of
Ref.~\cite{Whi05} and derive an optimal phase-space basis
for parallelogram bands, which gives the correct Ehrenfest time. 
This is done in a four step process.

{\bf Step [i].} Pick
a scattering band with phase-space area larger than $2\pi\hbar$, and cover
it with a lattice of PS-states. Both the lattice and the
states must be stretched and rotated to the same angle and aspect ratio as 
the band, and positioned in such a way as to minimize the number of
edge-of-band states. This is illustrated in Fig.~\ref{fig:band-and-ps-states}.
This can be done without relaxing either the mutual orthogonality,
or the normalization of the PS-states. 
Since the PS-states have the same aspect ratio as the band, their
longitudinal and transversal extensions
$x$ and $y$, as indicated in Fig.~\ref{fig:band-and-ps-states},
are given by ($\tau$ is the dwell time of
the band under consideration)\cite{footnote:lambda}
\begin{eqnarray}
x \simeq 
\left( {2\pi\hbar_{\rm eff}W_{\rm L}/W_{\rm R}} \right)^{1/2}
\exp [\lambda \tau/2] \;\; ; \;\; y \simeq 2 \pi \hbar_{\rm eff}/x.
\label{eq:aspect-ratio}
\end{eqnarray}
While we pay attention to minimizing their 
number, we do not include edge-of-band states in the 
basis at this stage. We will deal with them later in step [iv].

{\bf Step [ii].} We evolve the states generated in step [i] under the cavity's
dynamics. The lattice of PS-states
on an injection band uniquely determines the lattice of PS-states on
the exit band. All trajectories in the band under consideration
exit the cavity after a time shorter than the open cavity Ehrenfest time.
On this time scale the quantum dynamics of the PS-states
are well approximated by the Liouville flow \cite{Heller}
(see also Appendix \ref{appendix:evol-gaussian-wavepackets}).
This is a well-known property of coherent states 
that can be extended to the exponentially localized
PS-states that we consider here.
Thus a PS-state with initial spread of $\Delta Q$ in the 
unstable direction at $\tau=0$, will evolve into
a PS-state with spread $\Delta Q'\sim \Delta Q \e^{\lambda \tau}$ 
in the unstable direction at time $\tau$.
The initial spread in the stable direction 
is $\Delta P \simeq \hbar/\Delta Q$, and since the quantum dynamics
is Liouvillian inside classical bands, it is area preserving, i.e.
$\Delta P'\simeq \hbar/\Delta Q'$. 
The PS-states are stretched and rotated 
in the same manner as the exiting band, while still 
forming an orthonormal basis on that band. The orientation and stretch of the 
lattice and PS-states are
given in linear approximation (which eventually become valid as
$\hbar_{\rm eff} \to 0$)
by the eigenvalues and eigenvectors of the stability matrix
of the classical dynamics,
and thus have the same aspect ratio 
and angle as the exiting band. We choose to use these states to cover
that band in the phase-space. 

{\bf Step [iii].} We simply repeat the process in steps [i-ii] for each band 
with area $> 2 \pi \hbar$
not yet covered by PS-states.  The crucial point here is that bands cannot 
overlap; hence if we only place PS-states within classical
bands, Liouville blocking ensures that 
PS-states in different 
bands must be orthogonal with exponential accuracy.

{\bf Step [iv].} Steps [i-iii] generate an incomplete orthonormal basis
in the vector-space of lead modes. This basis can be made complete
by adding the adequate number of states orthogonal to 
those generated in steps [i-iii].  
This construction gives us very little information about the nature of these 
additional states, except that they must sit on more than one band.
They thus evolve in a quantum, stochastic manner,
and we refer to them as quantum PS-states.
The set of quantum PS-states divides into two broad categories: firstly those
which sit on many classical bands, secondly those which sit mostly,
but not completely on a single band. We already mentioned the second 
category of edge-of-band PS-states (the dashed ellipses in 
Fig.~\ref{fig:band-and-ps-states}). 
In Appendix ~\ref{sect:appendix-eob-states}
we estimate that the number of edge-of-band PS-state is
\begin{eqnarray}
N_{\rm eob} \simeq  (\lambda \tau_{\rm D})^{-1} N_{\rm qm} 
\label{eq:N_eob}
\end{eqnarray}
Hence they form a small fraction of the total number of quantum modes,
and we do not consider them separately from pure quantum modes.

\section{Scattering matrix in the phase-space basis}
\label{sect:S-and-T-in-ps-basis}

\subsection{Splitting of the scattering matrix and deterministic transmission}
\label{subsect:S-T-in-ps-basis}
By construction, the PS-basis is chosen so that there is a 
one-to-one correspondence between incoming and outgoing classical
PS-states,
given that their quantum dynamics can be approximated by the Liouvillian flow.
The unitarity of the scattering matrix means that the quantum 
PS-states remain orthogonal to the classical ones as they evolve inside the 
cavity. Thus, despite the fact that quantum PS-states
are not well described by the Liouvillian flow,  
they cannot penetrate the regions of phase-space containing
bands larger than $2\pi\hbar_{\rm eff}$.
In the PS-basis, each incoming classical PS-state goes to exactly
one outgoing classical PS-state, while each 
incoming quantum PS-state goes to multiple outgoing quantum PS-states, 
but no outgoing classical PS-states.
Correspondingly, the scattering matrix in the PS-basis
is of the form
\begin{eqnarray}
{\cal S} = {\cal S}_{\rm cl} \oplus {\cal S}_{\rm qm} 
=\left( \begin{array}{cc} 
{\cal S}_{\rm cl} & 0 \\ 0 & {\cal S}_{\rm qm}
\end{array}\right).
\label{eq:S-block-diag}
\end{eqnarray}
The matrix ${\cal S}_{\rm cl}$ is $N^{\rm cl} \times N^{\rm cl}$
while the matrix ${\cal S}_{\rm qm}$ is $N^{\rm qm} \times N^{\rm qm}$,
with $N^{\rm cl} = N^{\rm cl}_{\rm L} + N^{\rm cl}_{\rm R}$
and  $N^{\rm qm} = N^{\rm qm}_{\rm L} + N^{\rm qm}_{\rm R}$.

The matrix ${\cal S}_{\rm cl}$ has only one non-zero element in
each row and each column. The modes on the left and right of ${\cal S}$
can be reordered such that the transmission part 
${\bf t}_{\rm cl}$ of ${\cal S}_{\rm cl}$ is diagonal with all its non-zero
elements in the first $n$ elements of its diagonal, where $n$ is the
number of classical transmission modes.
Thus we can write
\begin{eqnarray}
{\bf t}_{\rm cl} 
= \left( \begin{array}{cc} 
\tilde{\bf t}_{\rm cl} & 0 \\ 0 & 0
\end{array} \right),
\label{eq:t_cl}
\end{eqnarray}
where all non zero-elements of ${\bf t}_{\rm cl}$ are contained
in the $n \times n$ matrix $\tilde{\bf t}_{\rm cl}$.
Doing the same for ${\bf t}'_{\rm cl}$, ${\bf r}_{\rm cl}$
and ${\bf r}'_{\rm cl}$, we write the 
classical part of the scattering matrix as
\begin{eqnarray}
{\cal S}_{\rm cl} 
\ \equiv\ 
\left( \begin{array}{cc} 
{\bf r}_{\rm cl} & {\bf t}'_{\rm cl} \\ {\bf t}_{\rm cl} & {\bf r}'_{\rm cl}
\end{array}\right)
\ =\ 
\left( \begin{array}{cccc} 
0 & 0 & \tilde{\bf t}'_{\rm cl}  & 0 \\ 
0 & \tilde{\bf r}_{\rm cl} & 0 & 0 \\
\tilde{\bf t}_{\rm cl} & 0 & 0 & 0 \\
0 & 0 & 0 & \tilde{\bf r}'_{\rm cl}
\end{array}\right) \; ,
\end{eqnarray}
where
$\tilde{\bf t}_{\rm cl}$ and 
$\tilde{\bf t}'_{\rm cl}$ are $n\times n$ matrices,
$\tilde{\bf r}_{\rm cl}$ is an 
$(N_L^{\rm cl}-n)\times(N_L^{\rm cl}-n)$ matrix 
and 
$\tilde{\bf r}'_{\rm cl}$ is an 
$(N_R^{\rm cl}-n)\times(N_R^{\rm cl}-n)$ matrix. 
The matrix $\tilde{\bf t}_{\rm cl}$ is diagonal with elements given by
\begin{eqnarray}
(\tilde{\bf t}_{\rm cl})_{ij} = \e^{\rmi \Phi_i}\delta_{ij} \; .
\label{eq:diag_tcl}
\end{eqnarray}
The matrix $\tilde{\bf r}_{\rm cl}$ also has
exactly one non-zero element in each row and each column. Its elements obey
\begin{eqnarray}
|(\tilde{\bf r}_{\rm cl})_{ij}| = |(\tilde{\bf r}_{\rm cl})_{ji}| 
= \left\{ \begin{array}{ccc}  
1 & \null & \hbox{ when $i$ reflects to $j$}, \\
0 & \null & \hbox{ otherwise}. \\
\end{array} \right.
\label{eq:symmetry-of-r_cl}
\end{eqnarray}
Anticipating discussions to come, we note that coherent backscattering
is carried here by matrix elements  
$(\tilde{\bf r}_{\rm cl})_{ij}$ where the $j$ denotes the time-reversal
of $i$. Since the classical probability to go from $i$ to its 
time-reversal is equal to the probability to go from $i$ to itself, the
number of nonzero such matrix elements is 
determined by the probability to sit on a periodic 
orbit which does not visit
the contact to any of the two leads except for one point.
This probability is zero. The absence of reflection
matrix elements carrying coherent backscattering 
in $\tilde{\bf r}_{\rm cl}$ qualitatively explains the
exponential suppression of coherent backscattering.

We next calculate the transmission matrix, ${\bf T} = {\bf t}^\dagger
{\bf t}$. The block diagonal nature of ${\cal S}$ in the PS-basis
given in Eq.~(\ref{eq:S-block-diag}),
ensures that ${\bf T}$ has the same structure in that basis,
hence
\begin{subequations}
\label{eq:T-block-diag} 
\begin{eqnarray}
{\bf T} &=&  {\bf T}_{\rm cl} \oplus {\bf T}_{\rm qm} 
=\left( \begin{array}{cc} 
{\bf T}_{\rm cl} & 0 \\ 0 & {\bf T}_{\rm qm}
\end{array}\right),\\
{\bf T}_{\rm cl} &=& {\bf t}_{\rm cl}^\dagger  {\bf t}_{\rm cl}
\;\;\; ; \;\;\;
{\bf T}_{\rm qm} = {\bf t}_{\rm qm}^\dagger  {\bf t}_{\rm qm} \ .
\end{eqnarray}
\end{subequations}
From Eq.~(\ref{eq:t_cl},\ref{eq:diag_tcl}) we 
get the eigenvalues of ${\bf T}_{\rm cl}$,
\begin{eqnarray}
T_i = \left\{ \begin{array}{ccc} 
1 & \qquad & \hbox{for } 1\leq i \leq n ,
\\ 0 & \qquad &  \hbox{for } n < i \leq N_{\rm L}^{\rm cl}.
\end{array}\right. 
\label{eq:T=0,1}
\end{eqnarray}
This is what we believe is the first proof
of a longstanding hypothesis, 
that in the classical limit the vast majority
of transmission eigenvalues are zero or one \cite{vanhouten}.
We know that there are $N_{\rm L}^{\rm cl}$ such classical modes,
with the remaining modes having a quantum nature, making them unlikely to
have transmission eigenvalues which are exactly zero or one.
The block-diagonal structure (\ref{eq:T-block-diag}) of the 
transmission matrix means that the dimensionless conductance, $g = \sum_i T_i$ 
and the Fano factor for shot noise of Eq.~(\ref{eq:fano}),
can be written as
\begin{eqnarray}
g&=&g_{\rm cl} + g_{\rm qm},
\\
F&=&
{g_{\rm cl} F_{\rm cl} + g_{\rm qm}F_{\rm qm} \over g_{\rm cl} + g_{\rm qm}}.
\end{eqnarray}
where we have introduced 
the conductance and Fano factor for the two cavities (classic and quantum),
$g_{\rm cl,qm} = \sum_{i \in {\rm cl,qm}} T_i$ and
$F_{\rm cl,qm} = [\sum_{i\in {\rm cl,qm}} T_i(1-T_i)]/[\sum_{i\in {\rm cl,qm}} 
T_i]$. From Eq.~(\ref{eq:T=0,1}) we see that
\begin{eqnarray}
g_{\rm cl} = n  
\qquad ; \qquad
F_{\rm cl} = 0.
\end{eqnarray}
Anticipating the calculation of
the average values $\langle g_{\rm cl,qm} \rangle$ (see next subsection),
\begin{equation}
F = F_{\rm qm} \exp[-\tELR/\tD].
\end{equation}
Since $F_{\rm qm} < 1$, one see there is an exponential suppression of $F$.

The PS-basis does not give us much information about the quantum PS-states.
However since
each incoming PS-state sits on multiple bands, exiting at different times,
it must couple to multiple outgoing PS-states.
It is extremely rare for these outgoing PS-states to 
be all transmitting (or all reflecting), and thus
we expect that the vast majority of
their transmission eigenvalues lie between
zero and one, and thus contribute to shot-noise.
However the fact that quantum and classical PS-states 
exist in two separate sub-blocks of the scattering 
and transmission matrices,
see Eqs.~(\ref{eq:S-block-diag}) and (\ref{eq:T-block-diag}), establishes the
two-phase fluid model \cite{Jac04}.

\subsection{The average Drude conductance}
\label{Drude}
  
From the estimates in Section \ref{sect:Ehren-times},
the ensemble-averaged Drude conductance is the sum of
the Drude conductances of the quantum and classical cavities,
\begin{subequations}
\begin{eqnarray}
\langle g \rangle_{\rm D} &=& \langle g_{\rm qm}\rangle + 
\langle g_{\rm cl}\rangle ,\\
\langle g_{\rm qm}\rangle 
&=& {N_{\rm L}^{\rm qm}N_{\rm R}^{\rm qm} \over 
N_{\rm L}^{\rm qm}+N_{\rm R}^{\rm qm}}
= {N_{\rm L}N_{\rm R} \over
N_{\rm L}+N_{\rm R}} \e^{-\tELR/\tD},
\\
\langle g_{\rm cl} \rangle
&=& {N_{\rm L}^{\rm cl}N_{\rm R}^{\rm cl} \over 
N_{\rm L}^{\rm cl}+N_{\rm R}^{\rm cl}} 
\nonumber\\
&=& {N_{\rm L}N_{\rm R} \over
N_{\rm L}+N_{\rm R}} [1-\e^{-\tELR/\tD}]. 
\label{eq:g_cl}
\end{eqnarray}
\end{subequations}
Thus the 
ensemble averaged Drude conductance is,
\begin{equation}\label{eq:drude}
\langle g \rangle_{\rm D} = N_{\rm L}N_{\rm R}/(N_{\rm L}+N_{\rm R}) .
\end{equation}  
The splitting of the cavity has little effect on 
$\langle g \rangle_{\rm D}$, even though
classical modes and quantum modes do not mix.
For strong asymmetry there is an additional term of
order $N_L(\lambda\tD)^{-2}$ on the right of Eq.~(\ref{eq:g_cl}),
however our calculation is not valid to that order 
because we ignored various order $(\lambda\tD)^{-1}$ terms,
such as edge-of-band states, in Eq.~(\ref{eq:Ncl_L}).

\subsection{Sample-to-sample conductance fluctuations}

The precise shape, size and number 
of the nonergodic phase-space structures fluctuates from sample
to sample. These fluctuations strongly affect $g_{\rm cl}$.
They are of a classical nature, and 
as such they induce the departure of conductance fluctuations from
their universal behavior \cite{TwoUCF,Jac04}.
Indeed, once $\hbar_{\rm eff}$ is small enough that quantum mechanics resolves
the largest scattering band on average, the sample-to-sample conductance  
fluctuations are dominated by the band fluctuations. Since each resolved
band carries a number $\propto \hbar_{\rm eff}^{-1}$ of channels, one 
expects sample-to-sample conductance fluctuations to exceed
the universal value in the semiclassical limit,
\begin{equation}\label{eq:magcf}
\sigma(g_{\rm cl}) \propto \hbar_{\rm eff}^{-1} \gg 1.
\end{equation}
The above argument predicts the onset for deviations
of $\sigma(g)$ from its universal behavior once the largest
band is quantum mechanically resolved, i.e. for 
$\hbar_{\rm eff} < (W/L)^2 \exp[-\lambda \tau_0]$, where $\tau_0$ is the
minimal escape time, of order the time of flight through the cavity. Both this
onset and the magnitude (\ref{eq:magcf}) of the sample-to-sample conductance
fluctuations have been observed numerically \cite{TwoUCF,Jac04}.

\section{Weak Localization}
\label{sect:weak-loc}

We calculate the leading order quantum correction to the 
Drude conductance. 
Our treatment applies both to the universal ($\tE/\tD \rightarrow 0$)
and deep semiclassical (finite $\tE/\tD$) regimes.
We present an explicit treatment of the
coherent backscattering peak showing that our theory preserves the unitarity
of the scattering matrix, as well as 
a calculation of the magnetoconductance.
Thus far we have constructed a special
basis (the ps-basis) which is aligned along the band structures
in the classical phase-space.
This made it easy to calculate the properties of the parts of phase-space
covered by bands larger than $2\pi\hbar$, 
allowing us to calculate the deterministic transmission eigenvalues $T=0,1$.
However the complexity of the quantum modes of that basis,
make it difficult to explicitly calculate the transport properties of those
modes \cite{wj-fano}. We therefore 
return to the lead-mode basis to calculate weak localization.
Unlike previous works we do not neglect the classical bands, however.
Indeed, our semiclassical approach is able to 
extract the conductance (including weak localization) of 
both the classical and quantum cavities. 

\begin{figure}
\includegraphics[width=7.5cm]{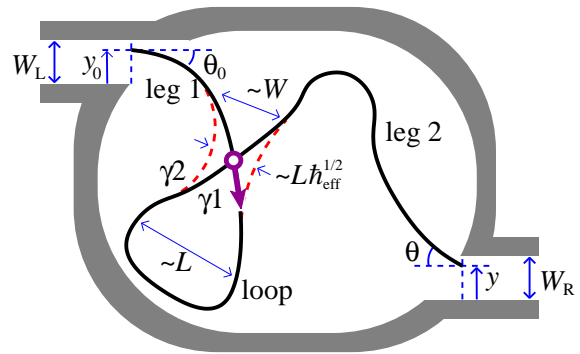}
\caption{\label{fig:rs-pair} (Color online) 
Sketch of the leading order quantum correction to the conductance.
Trajectory $\gamma1$ (solid line) 
is injected at ${\bf Y}_0=(y_0,\theta_0)$,
crosses itself and escapes at ${\bf Y}=(y,\theta)$.
Its first visit to the crossing (the open dot) occurs at 
${\bf R}_1=({\bf r}_1,\phi_1)$,
where ${\bf r}_1$ is the position in the cavity, and 
$\phi_1$ is the angle of the momentum to a reference axis 
(not shown).
Trajectory $\gamma2$ (dashed line)
starts and ends at the same positions as $\gamma1$,
however it avoids the crossing.
We divide $\gamma1$ into three segments; leg1, loop and leg2.
}
\end{figure}

\subsection{Drude conductance}
Semiclassically, the transmission matrix reads \cite{Bar93,fisher_lee},
\begin{eqnarray}\label{semicl-tr}
t_{ji} 
&=&
-(2\pi \rmi \hbar)^{-1/2}
\!\int_{\rm L} \! \! \rmd y_0 \int_{\rm R} \! \rmd y 
\sum_\gamma (\rmd p_y/\rmd y_0)^{1/2}_\gamma
\nonumber \\
& &\qquad \times \langle j|y\rangle  
\langle y_0| i \rangle 
\exp[\rmi S_\gamma /\hbar + \rmi \pi \mu_\gamma /2]
\, ,\qquad
\end{eqnarray} 
where $|i\rangle$ is the transverse wavefunction
of the $i$th lead mode. This expression 
sums over all  trajectories $\gamma$ (with classical action $S_{\gamma}$ 
and Maslov index $\mu_\gamma$) starting at $y_0$ on a cross-section of
the injection (L) lead and ending at $y$ on the
exit (R) lead. Inserting Eq.~(\ref{semicl-tr}) in the Landauer-B\"uttiker
formula for the 
conductance $g={\rm Tr} [{\bf t}^\dagger {\bf t}]$, one gets
a double sum over  trajectories, 
$\gamma1$ and $\gamma2$ and over lead modes, $|n\rangle$ and  $|m\rangle$.
We make the semiclassical approximation that
$\sum_n\langle y'|n\rangle\langle n|y\rangle \simeq \delta (y'-y)$ 
\cite{footnote:delta_hbar}.
The conductance is then given by a double sum over  trajectories 
which both go from $y_0$ on lead L to $y$ on lead R,
\begin{eqnarray}
{\rm Tr} [{\bf t}^\dagger {\bf t}]
&=& 
{1\over (2\pi \hbar)}
\!\int_{\rm L} \! \! \rmd y_0 \int_{\rm R} \! \rmd y   
\sum_{\gamma1,\gamma2} 
A_{\gamma1}A_{\gamma2}
\e^{\rmi\delta S/\hbar} \,.
\label{eq:conductance}
\end{eqnarray}
Here, $A_{\gamma}=[{\rm d} p_y/{\rm d}y_0]_\gamma^{1/2}$.
Reflection, $R={\rm Tr} [{\bf r}^\dagger {\bf r}]$, is given by the same 
expression, with both $y_0$ and $y$ on lead L.
We are interested in quantities 
averaged over variations in the energy or the cavity shape. 
For most $\{\gamma1,\gamma2\}$ 
the phase of a given contribution, $\delta S/\hbar$,
will oscillate
wildly with these variations, so the contribution 
averages to zero. The most obvious contributions that survive averaging
are the diagonal ones with $\gamma1=\gamma2$. These contributions give
the Drude conductance (\ref{eq:drude}).

We define $P({\bf Y},{\bf Y}_0;t)\de y\de \theta \de t$
as the product of the initial momentum along the injection lead,
$p_{\rm F}\cos \theta_0$, and the classical probability to go
from an initial position and momentum angle
${\bf Y}_0=(y_0,\theta_0)$ to within $(\de y,\de \theta)$ of
${\bf Y}=(y,\theta)$ in a time within $\de t$ of $t$. 
Then the sum over
all  trajectories $\gamma$ from  $y_0$ to $y$ is
\begin{eqnarray}
\sum_\gamma \!
A^2_\gamma \;
\! [\cdots]_\gamma 
\!\! &=& \!\!
\int_0^\infty 
\! \rmd t \; \int_{-\pi/2}^{\pi/2} \rmd \theta_0 \; 
\int_{-\pi/2}^{\pi/2} \rmd \theta \; 
\nonumber \\
&& \times \; P({\bf Y},{\bf Y}_0;t) 
\; [\cdots]_{{\bf Y}_0}.
\label{eq:gamma-sum-to-Pintegral}
\end{eqnarray}
For an individual system, $P$ has $\de$-functions for
all classical trajectories.  However averaging over 
an ensemble of systems or over energy 
gives a smooth function
\begin{eqnarray}
\langle P({\bf Y},{\bf Y}_0;t) \rangle = \frac{
p_{\rm F} \cos \theta_0 \cos \theta }
{2 (W_{\rm L}+W_{\rm R})\tau_{\rm D}} \; 
\exp[-t/\tau_{\rm D}] \, .
\label{eq:average-P}
\end{eqnarray}
This latter expression (\ref{eq:average-P}) is valid as long as no
restriction is imposed on the trajectory inside the cavity.
Using Eqs.~(\ref{eq:gamma-sum-to-Pintegral}) and (\ref{eq:average-P}) 
to calculate the conductance within the diagonal approximation,
one recovers the Drude conductance (\ref{eq:drude}),
\begin{subequations}
\begin{eqnarray}
g_{\rm diag} = g_{\rm D} &=& {\NL\NR \over \NL+\NR} \, ,
\\
R_{\rm diag} &=&{\NL^2 \over \NL+\NR}\,.
\end{eqnarray}
\end{subequations}
One also sees that at the level of the diagonal approximation, there
is unitarity. We stress that, unlike in Ref.~\cite{Ric02},
we do not include coherent backscattering in the 
diagonal contribution, it is dealt with separately below.

\subsection{Weak localization for transmission}

\begin{figure}
\centerline{\hbox{\includegraphics[width=8.5cm]{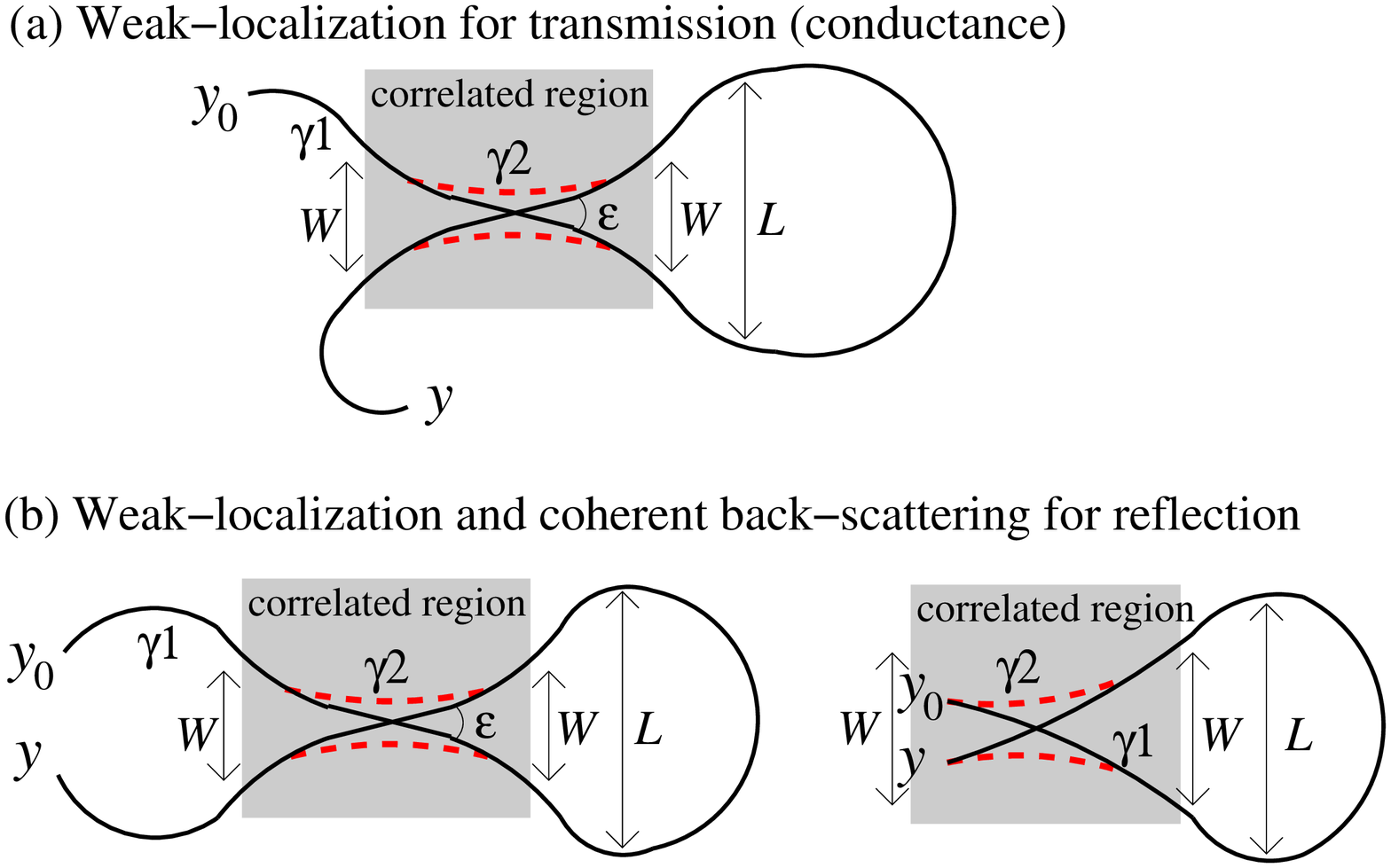}}}
\caption[]{\label{fig10} (Color online) 
Sketches of the trajectory pairing which give the leading off-diagonal
contributions to (a) transmission (conductance) and (b) reflection.
All contributions involve a trajectory $\gamma1$ crossing itself at an
angle $\epsilon$, and a trajectory
$\gamma2$ which avoids the crossing. The action difference
between the two trajectories is thus small and does not fluctuate under 
averaging. 
For transmission, $y_0$ is on L lead and $y$ is on R lead,
for reflection both $y_0$ and $y$ are on L lead.
There are two reflection contributions.
On the left is weak localization, and on the right is
coherent backscattering (details of the latter are in Fig.~\ref{fig12}).
}
\end{figure}

A pair of trajectories giving the leading order
correction to the Drude conductance is shown in Fig.~\ref{fig:rs-pair}.
The  trajectories are paired almost everywhere except in the
vicinity of an encounter \cite{Ric02}. 
Going through an encounter, one of the  trajectories intersects itself,
while the other one avoids the crossing. Thus, they travel along the
loop they form in opposite direction. It has been shown in Ref.~\cite{Ric01}
that for any self-intersecting trajectory with a 
small enough crossing angle $\epsilon$, there exists a partner,
crossing-avoiding outer trajectory. For the relevant case
of small $\epsilon$, the probability to find a weak localization
pair is thus given by the probability to find
a self-intersecting trajectory.
The two  trajectories are always close enough to each other
that their stability is the same, i.e. one can set
$\sum_{\gamma1,\gamma2} A_{\gamma1}A_{\gamma2}
\rightarrow \sum_{\gamma1} A_{\gamma1}^2$.
To evaluate the weak localization correction to conductance, we 
perform a calculation similar to Ref.~\cite{Ric02}, 
adding the crucial fact that pair of  trajectories such as depicted in 
Fig.~\ref{fig:rs-pair} have highly correlated escape probabilities 
due to the presence of an encounter \cite{Bro05}. The situation
is depicted in more detail in Fig.~\ref{fig10}. 

The presence of
the encounter introduces two new ingredients, both of these were
overlooked in Ref.~\cite{Ric02}.
First, pairs of  trajectories leaving an encounter escape the cavity in either
a correlated or an uncorrelated way. Uncorrelated escape occurs
when the perpendicular distance between the  trajectories 
is larger than the width $W_{\rm L,R}$ of the leads. This
requires a minimal time
$T_W(\eps)/2$ between encounter and escape, where \cite{caveat2}
\begin{eqnarray}
T_W(\eps) = \lambda^{-1} \ln [\eps^{-2} (W/L)^2] \ .
\end{eqnarray}  
The two pairs of  trajectories
then escape in an uncorrelated manner, typically at completely different
times, with completely different momenta (and possibly through 
different leads). Correlated escape occurs in the other situation
when the distance between the  trajectories at the time of escape is less 
than $W_{\rm L,R}$.
Then the two pairs of  trajectories escape together, at the same
time through the same lead. This latter process affects coherent
backscattering (see Fig.~\ref{fig10}).
The second new ingredient 
is that the survival probability for a trajectory
with an encounter is larger than that of a generic trajectory. This is so,
because the encounter duration affects the escape probability only once. 
In other words, if the trajectory did not escape in its first passage through 
the encounter, neither will it during its second passage (this was first noticed
in Ref.~\cite{Bro05}). 

\begin{figure}
\includegraphics[width=7cm]{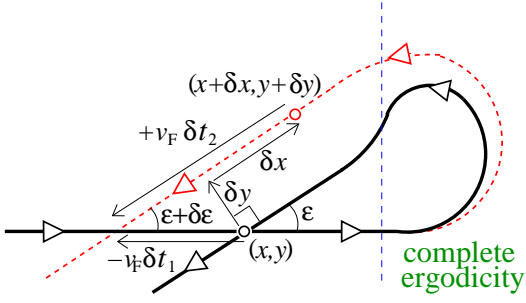}
\caption{\label{fig:ergod-cross}
(Color online) Sketch of a  trajectory (solid line)
which crosses itself at point ${\bf r}_1=(x,y)$, visiting
this point first at time $t_1$ and second at time $t_2$.
Superimposed is an infinitesimally different trajectory (dashed line)
which also visits the point 
$(x,y)$ at $t_1$, but is at point 
${\bf r}_2 =(x+\delta x,y+\delta y)$ at time $t_2$.
This trajectory also intersects itself;
however, it visits the self-intersection (which is no longer $(x,y)$) 
at times $t_1+\delta t_1$ and $t_2+\delta t_2$.
}
\end{figure}

We calculate the contribution from pairs of transmitting  trajectories 
sketched in
Fig.~\ref{fig10}a. For preserved time-reversal symmetry,  
the action difference for this contribution is 
$\delta S_{\rm wl}= E_{\rm F}\eps^2/\lambda$ \cite{Ric02}.
We now note that the probability to go from ${\bf Y}_0$ to 
        ${\bf Y}$ in time $t$, is the product of the probability to go from
        ${\bf Y}_0$ to a point on the energy surface inside the cavity 
        ${\bf R}_1 = ({\bf r_1},\phi_1)$ (where $\phi_1$ defines 
        the direction of the momentum) in time $t_1$ and the probability to go 
        from ${\bf R}_1$ to ${\bf Y}$ in time $t-t_1$, when one integrates 
        ${\bf R}_1$ over the energy surface ${\cal C}$. 
        Thus the quantity $P$ introduced above can be written as
\begin{eqnarray}
P({\bf Y},{\bf Y}_0;t)
\!&=& \!\!
\int_{\cal C} \rmd {\bf R}_2 \rmd {\bf R}_1
\tilde{P}({\bf Y},{\bf R}_2;t-t_2)
\nonumber \\
& \times &
\tilde{P}({\bf R}_2,{\bf R}_1;t_2-t_1)
P({\bf R}_1,{\bf Y}_0;t_1) \, . \qquad
\end{eqnarray} 
         where $\tilde{P}({\bf R}_2,{\bf R}_1;t)$ is the probability 
         density to go from ${\bf R}_1$ to ${\bf R}_2$ in time $t$, 
         but $P({\bf R_1},{\bf Y}_0;t)$ is a probability density multiplied 
         by the injection momentum, $p_{\rm F} \cos \theta_0$.
We then restrict the probabilities inside the
integral to  trajectories which cross themselves
at phase-space positions ${\bf R}_{1,2}$
with the first (second) visit to the crossing occurring at
time $t_1$ ($t_2$). 
Using Fig.~\ref{fig:ergod-cross}, we write $\rmd {\bf R}_2 = v_{\rm F}^2 
\sin \eps \rmd t_1 \rmd t_2 \rmd \eps$
and set ${\bf R}_2\equiv({\bf r}_2,\phi_2)=({\bf r}_1,\phi_1\pm \eps)$.
Next, we note that the duration of the loop must exceed
$T_L(\eps) = \lambda^{-1} \ln [\eps^{-2}]$,
because for shorter times, two trajectories leaving an encounter
remain close enough to each other that their relative dynamics is
hyperbolic, 
and the probability of forming a loop is zero.
   Similarly the path cannot transmit unless $t_1, t-t_2 > T_W(\eps)$,
   because for  $t_1, t-t_2 < T_W(\eps)$ the legs (see Fig.~\ref{fig:rs-pair})
   are so close to each other
   that if one leg escapes through a given lead the other one will escape 
   with it through the {\it same} lead.
Thus the probability that a  trajectory starting at ${\bf Y}_0$
crosses itself  
at an angle $\pm\eps$
and then transmits, multiplied by its injection momentum 
$p_{\rm F}\cos \theta_0$, is 
\begin{eqnarray}
I( {\bf Y}_0,\eps)
\! &=& \! 
2v_{\rm F}^2 \; \sin \eps \;
\int_{T_L+T_W}^\infty \rmd t 
\int_{T_L+T_W/2}^{t-T_W/2} \! \rmd t_2
\int_{T_W/2}^{t_2-T_L} \! \rmd t_1
\nonumber\\
& & \times 
\int_{\rm R} \rmd {\bf Y} \int_{\cal C} \rmd {\bf R}_1 \;
\tilde{P}({\bf Y},{\bf R}_2;t-t_2) 
\nonumber \\[2mm]
& & \times \,\tilde{P}({\bf R}_2,{\bf R}_1;t_2-t_1) \; P({\bf R}_1,{\bf Y}_0;t_1)  \, ,
\quad \quad 
\label{eq:I}
\end{eqnarray}
where $T_W, T_L$ are shorthand for $T_W(\eps),T_L(\eps)$.

To get the weak localization correction to conductance
we sum only contributions where $\gamma1$ crosses itself,
we then take twice the real part of this result to include the contributions
where $\gamma1$ avoids the crossing (and hence $\gamma2$ crosses itself).
Thus 
\begin{eqnarray}\label{gwl}
g_{\rm wl}
&=& 
(\pi \hbar)^{-1}  
\int_{\rm L} \rmd {\bf Y}_0 \rmd \eps 
{\rm Re}\big[\e^{\rmi \de S_{\rm wl}/\hbar}\big] 
\big\langle I( {\bf Y}_0,\eps) \big\rangle .\qquad
\end{eqnarray}
We perform the average of the $P$s as follows.
Within $T_W(\eps)/2$ of the crossing the two legs of a
self-intersecting  trajectory are so close
to each other that their joint escape probability is 
the same as for a single  trajectory. 
Self-intersecting trajectories thus
have an enhanced survival probability compared to non-crossing trajectories
of the same length, i.e. 
the duration of the crossing
must be counted only once in the survival probability \cite{Bro05}.
Outside the correlated region, the legs can escape
independently through either lead at anytime. Furthermore,
the probability density for the  trajectory 
going to a given point in phase-space
is assumed to be uniform.  
  Thus the probability density for leg 1 gives
  $\langle P({\bf R}_1,{\bf Y}_0;t_1)\rangle 
  = (2\pi A)^{-1} \exp (-t_1/\tau_{\rm D}) \times p_{\rm F}\cos \theta_0$,
  and the loop's probability density is
  $\langle \tilde{P}({\bf R}_2,{\bf R}_1;t_2-t_1)\rangle 
  = (2\pi A)^{-1} \exp \{-[t_2-t_1-T_{\rm W}(\eps)/2]/\tau_{\rm D}\}$
  (Ref.~\cite{footnote:mea-culpa}). 
  Finally the {\it conditional} probability density for leg 2 
  (given that leg 1 exists for a time $t_1> T_W(\eps)$) is 
  $\langle \tilde{P}({\bf Y},{\bf R}_2;t-t_2)\rangle 
  = [2(W_L+W_R)\tau_{\rm D}]^{-1} \cos \theta
  \exp \{-[t-t_2-T_{\rm W}(\eps)/2]/\tau_{\rm D}\}$.
Hence one finds
\begin{eqnarray}
& & \hskip -5mm
\langle \tilde{P}({\bf Y},{\bf R}_2;t-t_2) \; 
\tilde{P}({\bf R}_2,{\bf R}_1;t_2-t_1) \; 
P({\bf R}_1,{\bf Y}_0;t_1) \rangle  \nonumber \\
&=&\frac{1}{(2 \pi A)^2} \; \frac{p_{\rm F} \cos \theta \cos \theta_0}{2 (\WL+\WR) \tD} \; \exp[-(t-T_L(\epsilon))/\tD] , \quad \quad  
\end{eqnarray}
so that $\langle I( {\bf Y}_0,\eps) \rangle$ becomes 
\begin{eqnarray}
\big\langle I({\bf Y}_0,\eps) \big\rangle
&=&{(v_{\rm F} \tau_{\rm D})^2 \over \pi A} \; p_{\rm F} \sin \epsilon \cos \theta_0 \; \nonumber \\
&& \times {N_{\rm R} \over N_{\rm L}+N_{\rm R}}
\exp[-T_L(\eps)/\tau_{\rm D}] \, .
\end{eqnarray}
We insert this into Eq.~(\ref{gwl}). The $\epsilon$-integral 
is dominated by contributions with $\epsilon \ll 1$, so that we
write $\sin \eps \simeq \eps$ and push the upper bound for the
$\epsilon$-integration to infinity. The $\eps$-integral can
then be computed to give
an Euler $\Gamma$-function \cite{Ada03}. To leading order in 
$(\lambda \tD)^{-1}$ it equals
$-\pi \hbar (2E_{\rm F} \tau_{\rm D})^{-1}$.  
The integral over ${\bf Y}_0$ yields a factor of $2 \WL$. 
Finally noting that $\NL=(\pi \hbar)^{-1}p_{\rm F}\WL$
and $(\NL+\NR)^{-1}= (mA)^{-1}\hbar \tau_{\rm D}$, the weak localization 
correction to the conductance reads
\begin{eqnarray}
g_{\rm wl} &=& -{\NL \NR \over (\NL+\NR)^2 } \exp[-\tEc/\tD] \, .
\label{eq:g_wl}
\end{eqnarray}
We see that weak localization is exponentially suppressed with
$\tEc/\tD$ in term of the closed cavity Ehrenfest time 
$\tEc \equiv \lambda^{-1}[\hbar_{\rm eff}^{-1}]$.
\begin{figure}
\centerline{\hbox{\includegraphics[width=8.5cm]{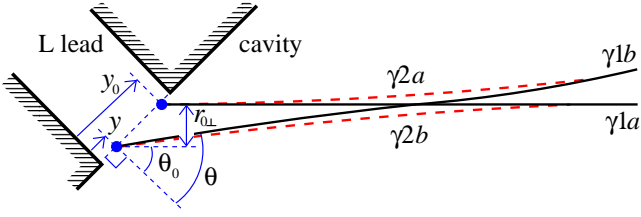}}}
\caption[]{\label{fig12} (Color online) 
Trajectories for the backscattering contributions to reflection.
Trajectory $\gamma 1$ (solid black line)
start on the cross-section of the L lead at position
$y_0$ with momentum angle $\theta_0$ and
ends at $y$ with momentum angle $\theta$.
In the basis parallel and perpendicular to $\gamma1$ at injection, 
the initial position and momentum of path $\gamma1$ at exit are
$r_{0\perp} = (y_0-y)\cos \theta_0$, $r_{0\pll} = (y_0-y)\sin \theta_0$  
and $p_{0\perp} \simeq -p_{\rm F} (\theta-\theta_0)$.
}
\end{figure}

\subsection{Quantum corrections to reflection}

The above result (\ref{eq:g_wl}) has already been derived in 
Ref.~\cite{Ada03} by a similar approach. We go beyond that
by showing explicitly that our semiclassics preserves the unitarity
of the scattering matrix. 
There are two leading-order off-diagonal corrections to reflection. 
They are shown in Fig.~\ref{fig10}(b). The first is weak localization
while the second is coherent backscattering.
The former reduces the probability of reflection to arbitrary
momentum, while the latter enhances the probability of reflection
to the time-reverse of the injection state.
The distinction between these two contributions
is whether or not the legs escape while correlated:
for weak localization the legs escape independently, while for
coherent backscattering the legs escape together in a 
correlated manner.

\subsubsection{Weak localization}

The weak localization contribution to reflection, $R_{\rm wl}$,
is derived in the same manner as $g_{\rm wl}$, replacing however
a factor of $\WR/(\WL+\WR)$ with $\WL/(\WL+\WR)$. One obtains
\begin{eqnarray}
R_{\rm wl} = -{\NL^2 \over (\NL+\NR)^2} \exp[-\tEc/\tD]
\label{eq:R_wl}
\end{eqnarray}

\subsubsection{Coherent backscattering}

Contributions to coherent backscattering are shown in Fig.~\ref{fig10}b,
with Fig.~\ref{fig12} showing the trajectories in the correlated region 
in more detail. These contributions require special care because
(i) their action phase difference $\delta S_{\rm cbs}$ is not given
by the Richter-Sieber expression used so far and (ii) injection and exit
positions and momenta are correlated.

From Fig.~\ref{fig12}, and noting 
that $\gamma2b$ decays exponentially towards $\gamma1a$, we find
the action difference between these two path segments to be
\begin{eqnarray}
S_{2b}-S_{1a} &=& p_{\rm F}(y_0-y)\sin \theta_0
\nonumber \\ 
& & \qquad + \half m\lambda (y_0-y)^2\cos^2 \theta_0 \, . \quad
\label{eq:correlated-action-diff}
\end{eqnarray}
We have dropped cubic terms which only give $\hbar$-corrections 
to the stationary-phase integral. 
The action difference between $\gamma2a$ and $\gamma1b$ 
has the opposite sign for $y_0-y$
and $\theta_0$ replaced by $\theta$.
We get for the total action difference, 
in terms of $(r_{0\perp},p_{0\perp})$,
\begin{eqnarray}
\de S_{\rm cbs} 
&=& -(p_{0\perp}+m\lambda r_{0\perp})r_{0\perp}\,.
\label{eq:deltaS_cbs}
\end{eqnarray}
The coherent backscattering contribution to the reflection reads
\begin{eqnarray}
R_{\rm cbs} 
&=& \! 
(2\pi \hbar)^{-1}
\!\int \!\! \rmd {\bf Y}_0  \rmd {\bf Y}
\int_0^\infty \!\! \rmd t 
\nonumber \\ 
& & \times
\langle P({\bf Y},{\bf Y}_0;t)  \rangle \;{\rm Re}\big[
\e^{\rmi \de S_{\rm cbs}/\hbar}\big] \, .
\label{eq:R_cbs}
\end{eqnarray}
    Note that this contains all those contributions where
    $\gamma1$ crosses itself and all those contributions 
    where it avoids crossing (so $\gamma2$ crosses itself), 
    thus there is no need to take twice the real part here 
    (unlike for $g_{\rm wl}$).
To perform the average 
we define $T'_{W}(r_{0\perp}, p_{0\perp})$ 
and $T'_{L}(r_{0\perp}, p_{0\perp})$ 
as the times for which the perpendicular distance between the
$\gamma1a$ and $\gamma1b$ is $W$ and $L$, respectively.
For times less than $T'_{W}(r_{0\perp}, p_{0\perp})$ 
the escape probability for two trajectories is the same as 
for one, while for times longer than this
the  trajectories evolve and escape independently.
For $R_{\rm cbs}$ we consider only those trajectories
that form a closed loop, however they cannot close until the two 
trajectory segments are of order $L$ apart. 
The $t$-integrals must have a lower cut-off at 
$2T'_L(r_{0\perp}, p_{0\perp})$,
hence 
\begin{eqnarray}
& &\int_{\rm R} \! \rmd {\bf Y}
\int_{2T'_L}^\infty \rmd t 
\langle P({\bf Y},{\bf Y}_0;t)\rangle 
\nonumber \\
& & \qquad 
= p_{\rm F} \cos \theta_0 \; \frac{N_{\rm L}}{N_{\rm L} +N_{\rm R}} 
\exp[-T'(r_{0\perp}, p_{0\perp})/\tau_{\rm D}] ,\  
\, \qquad
\end{eqnarray}
where $T' (r_{0\perp}, p_{0\perp})
= 2T'_L(r_{0\perp}, p_{0\perp})-T'_W(r_{0\perp}, p_{0\perp})$.
For small $(p_{0\perp}+ m\lambda r_{0\perp})$ we estimate
\begin{eqnarray}
T'(r_{0\perp}, p_{0\perp}) 
&\simeq& 
\lambda^{-1} \ln \left[
{W(p_{0\perp}+ m\lambda r_{0\perp}) \over m\lambda L^2 }\right]
\, . \qquad 
\end{eqnarray}
We substitute the above expression into $R_{\rm cbs}$,
write  
$p_{\rm F} \cos\theta_0 \rmd {\bf Y}_0 =  
\rmd y_0\rmd (p_{\rm F} \sin\theta_0 )
= \rmd r_{0\perp} \rmd p_{0\perp}$
\cite{Bar93}, and then make the substitution
$\tilde{p}_0=p_{0\perp}+m\lambda r_{0\perp}$.
We evaluate the $r_{0\perp}-$integral over a range of order $W_{\rm L}$,
take the limits on the resulting $\tilde{p}_0$-integral 
to infinity
and write it in terms of Euler $\Gamma$-functions.
Finally we systematically drop all terms ${\cal O}(1)$ inside
logarithms. The result is that
\begin{eqnarray}\label{eq:cbs}
R_{\rm cbs}
&=& \frac{\NL}{\NL+\NR} \exp[-\tEc/\tD] \,.
\end{eqnarray}
Thus we see that 
coherent backscattering is also suppressed exponentially in exactly
the same manner as weak localization.
Hence $R_{\rm cbs} + R_{\rm wl}= -g_{\rm wl}$ and unitarity is preserved.

\subsection{The off-diagonal nature of coherent backscattering}

Continuous families of  trajectories that are present in open chaotic systems,
such as $\gamma2$ and $\gamma1$ in Fig.~\ref{fig12},
have an action difference given in 
Eq.~(\ref{eq:deltaS_cbs}). This action difference does not fluctuate
under energy or sample averaging, moreover,
these contributions are not diagonal in the lead mode basis.
The stationary phase integral over such  trajectories is 
dominated by $p_{0\perp}\simeq -m\lambda r_{0\perp}$
where $r_{0\perp}$ is integrated over the total lead width.
Thus $p_{0\perp}$ varies over a range of order $ m\lambda W 
\gg \pi \hbar/ W$,
coupling to many lead modes. 
Such contributions were not taken into account in the previous
analysis of coherent backscattering \cite{Ric02}.   
This caused the erroneous belief 
(of the authors of the present article amongst others)
that coherent backscattering 
originates from  trajectories that return to any point in the L lead 
with $\theta\simeq \pm \theta_0$,
which would have implied that the coherent backscattering
was independent of the Ehrenfest time.

Once we correctly sum the many off-diagonal
contributions to $\sum_{nm} |r_{nm}|^2$ which have an encounter 
near the L lead, we conclude that coherent backscattering
approximately doubles the weight of all returning trajectories in a strip
defined by 
\begin{eqnarray}
\theta-\theta_0 \simeq - p_{\rm F}^{-1} m \lambda (y-y_0) \cos \theta_0 \ 
\end{eqnarray}
across the lead. 
This strip sits on the stable axis of the classical dynamics,
with a width in the unstable direction of order 
$\hbar(p_{\rm F}W)^{-1}$.
Therefore, trajectories in the strip first converge toward each other,
and only start diverging at a time of order $\tEo/2$. 
Such  trajectories cannot form a loop 
on times shorter than $\tEo+\tEc$.

\subsection{Magnetoconductance}

A weak magnetic field has very little effect on the classical dynamics.
Its dominant effect is to generate a phase difference between two trajectories 
that go the opposite way around a closed loop.
This phase difference is ${\cal A}_{\rm loop}\Phi$ where ${\cal A}_{\rm loop}$ 
is the directed area enclosed by the loop, 
and $\Phi$ is the flux in units of the flux quantum. 
To incorporate this in the theory we must
introduce a factor of $\exp [\rmi {\cal A}_{\rm loop}\Phi]$
into $I({\bf Y}_0,\eps)$ in Eq.~(\ref{eq:I})
and inside the average in $R_{\rm cbs}$ in Eq.~(\ref{eq:R_cbs}).  
To average
$\langle \exp [\rmi {\cal A}_{\rm loop}\Phi] \rangle$,
we divide the loop into two parts --- 
the correlated part (within $T_L(\eps)/2$
of the crossing), and the uncorrelated part (the rest of the  
loop). We average the two parts separately.

For the uncorrelated part, we use the fact that the distribution of
area enclosed by classical scattering trajectories in a chaotic system is
Gaussian with zero mean and a variance which increases linearly with time 
\cite{Bar93}. One then has,
\begin{eqnarray}\label{eq:phi-avg}
\langle e^{\rmi {\cal A}_{\rm uncorr}\Phi}\rangle
= \exp [- \al A^2\Phi^2 \, (t_2-t_1- T_L(\eps))/\tau_{\rm f}],
\,  \quad
\end{eqnarray}
where, $A$ is the area of the cavity, $\alpha$ is a system-dependent
parameter of order unity, and $\tau_{\rm f}$ is the time of flight
between two consecutive bounces at the cavity's wall.

We comment on the correlated part in Appendix \ref{sect:magneto-correl},
where we show that it provides at most only small corrections
${\cal O}(\tau_{\rm f}/\tau_{\rm D})$ which we henceforth
ignore. Multiplying the integrand in Eq.~(\ref{eq:I}) with (\ref{eq:phi-avg}),
and integrating over $t_1,t_2$ gives
\begin{eqnarray}
\big\langle I({\bf Y}_0,\eps) \big\rangle
&=&{(v_{\rm F} \tau_{\rm D})^2 \over \pi A} \; p_{\rm F} \sin \epsilon \cos \theta_0 \; \nonumber \\
&& \times {N_{\rm R} \over N_{\rm L}+N_{\rm R}}
{\exp[-T_L(\eps)/\tau_{\rm D}] \over 1 + \al A^2 (\tD/\tau_{\rm f})\Phi^2 } 
 \, .
\end{eqnarray}
After a similar analysis for $R_{\rm cbs}$,
we conclude that for finite flux, the quantum corrections
to the average conductance acquire a Lorentzian shape,\\
\begin{subequations}
\begin{eqnarray}\label{eq:gwl-lorentz}
g_{\rm wl}(\Phi)   &=& -{\NL \NR \over (\NL+\NR)^2 } \ {\exp[-\tEc/\tD] 
\over 1 + \al A^2 (\tD/\tau_{\rm f})\Phi^2 }, \\
R_{\rm wl}(\Phi)   &=& -{\NL^2 \over (\NL+\NR)^2} \ {\exp[-\tEc/\tD]
\over 1 + \al A^2 (\tD/\tau_{\rm f})\Phi^2 },  \\
R_{\rm cbs}(\Phi)  
&=& \ \frac{\NL}{\NL+\NR} \ {\exp[-\tEc/\tD] 
\over 1 + \al A^2 (\tD/\tau_{\rm f})\Phi^2 }. 
\end{eqnarray}
\end{subequations}
Interestingly enough, there is no Ehrenfest dependence in the width of the 
Lorentzian.

\subsection{Weak localization in the two-fluid model}

Weak localization can also be calculated in the framework of the
special basis constructed in the first half of this paper.
We can split all contributions to conductance into
classical and quantum contributions using the classical dynamics.
By construction the classical modes couple to the  trajectories
shorter than $\tEo$, while the quantum ones couple
to the ones longer than $\tEo$.  We cut the time-integrals
in all the above quantities at $\tEo$,
the result is that the classical cavity has
\begin{subequations}
\begin{eqnarray}
g_{\rm D}^{\rm cl} &=& {\NL\NR \over \NL+\NR} [1-\e^{-\tEo/\tD}],
\\
R_{\rm diag}^{\rm cl} &=& {\NL^2 \over \NL+\NR} [1-\e^{-\tEo/\tD}],
\\
g_{\rm wl}^{\rm cl} &=& R_{\rm wl}^{\rm cl} \,=\, R_{\rm cbs} \,=\, 0 .
\end{eqnarray}
\end{subequations}
The quantum modes carrying the remaining contributions. Inserting by hand
the phase-space splitting into the sum rules and bound of time integrations 
in the
above semiclassical treatment, one recovers the exponential suppression
of weak localization, Eq.~(\ref{eq:g_wl}).
The quantum fluid is thus clearly not RMT, which invalidates the
effective RMT model \cite{Sil03}.
This is because contributions to weak localization and coherent 
backscattering come from trajectories longer than $\tEo+\tEc$,
and the proportion of such trajectories in the quantum cavity
goes like $\exp[-\tEc/\tD]$. 

The existence of classical bands is key to both the existence of the
two separate fluids (block diagonal nature of ${\cal S}$, 
Eq.~(\ref{eq:S-block-diag}))
and to the exponential suppression of coherent backscattering 
and the unitarity of the semiclassical theory presented here
(see also \cite{wj-fano}).


\section{Numerical Simulations}
\label{sec:numerics}

We finally check our semiclassical theory for weak localization 
at finite $\tEc/\tD$ against numerical simulations.
We consider open systems with fully developed chaotic dynamics,
for which $\tD \gg 1$. Because $\tEc$ grows 
logarithmically with the Hilbert space size $M = \hbar_{\rm eff}^{-1}$, 
and since we want to investigate the regime of finite $\tEc/\tau_D$, 
we model the electron dynamics
by the kicked rotator map \cite{TwoSN,Jac04,Henning05}.
The Hamiltonian is given by
\begin{eqnarray}\label{krot}
H = \frac{(p+p_0)^2}{2} + K \cos(x+x_0) \sum_n \delta(t-n \tau_{\rm f}).
\end{eqnarray}
The kicking strength $K$ drives the dynamics from
integrable ($K=0$) to fully chaotic [$K\agt 7$, with Lyapunov exponent
$\lambda \tau_{\rm f}\approx\ln (K/2)$]. The parameters $p_0$ and $x_0$ are 
introduced to break the Hamiltonian's two symmetries \cite{izrailev}. 
Only when these two symmetries are broken does one witness a crossover
from the $\beta=1$ to the $\beta=2$ universality class \cite{Meh91}, 
corresponding to breaking the time reversal symmetry \cite{izrailev}. 
The procedure of varying $p_0$ followed in Refs.~\cite{TwoWL,Bro05}
results in a strongly non-Lorentzian magnetoconductance. 
Thus the agreement between numerics and analytics in Ref.~\cite{Bro05} 
could not be extended to the magnetoconductance curve,
which is the trademark of weak localization. 
This motivated us to perform numerical investigations 
following the same procedure as in Ref.~\cite{bardarson}, i.e. taking
a finite, constant $p_0$, while varying $x_0$. In this case, the 
magnetoconductance curves are Lorentzian.

\begin{figure}
\begin{center}
\hspace{-0.5cm}\includegraphics[width=7cm]{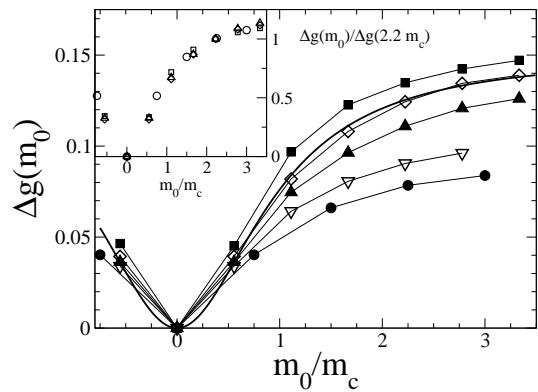}
\end{center}
\caption{\label{fignum1} Magnetoconductance curves 
$\Delta g(m_0)=g(m_0)-g(0)$ for the
open kicked rotator model (defined in the text) at fixed classical 
configuration $K=14$ ($\lambda \approx 1.95$), 
$\tD/\tau_{\rm f}=5$, and different Hilbert space sizes,
$M=256$ (squares, $\tEc/\tD \simeq 0.57$), $M=512$ (diamonds, 
$\tEc/\tD\approx 0.64$), 
$M=1024$ (upward triangles, $\tEc/\tD \approx 0.71$), 
$M=2048$ (downward triangles, 
$\tEc/\tD\approx 0.78$), and $M=4096$ (circles, $\tEc/\tD\approx 0.85$). 
The solid line gives the best Lorentzian fit for the $M=512$ curve,
$\Delta g(m_0)=0.15-0.15/[1+1.03 \; (m_0/m_c)^2]$.
Data have been obtained after averaging over 100000 (1000 
classically different samples, each with 100 different quasienergies
for $M=256$)
to 25000 (500 classically different samples, each with
50 different quasienergies for $M=4096$) different samples.
Inset: rescaled magnetoconductance data,
$\Delta g(m_0)/\Delta g(2.2 m_c)$, the data collapses onto a single curve, 
confirming out theory that $m_c$ does not depend on $\tEc$.}
\end{figure}

We consider a toroidal classical phase-space $x,p \in[0,2 \pi]$, 
and open the system by
defining contacts to ballistic leads via two absorbing phase-space strips
$[x_L-\delta x,x_L+\delta x]$ and $[x_R-\delta x,x_R+\delta x]$, 
each of them with a width $2 \delta x=\pi/\tD$.
We quantize the map by discretizing the 
momentum coordinates as $p_l=2 \pi l/M$, $l=1,\ldots M$. 
A quantum representation of the Hamiltonian (\ref{krot}) is
provided by the unitary $M \times M$ 
Floquet operator $U$, 
which gives the time evolution for one iteration of the map. 
For our specific choice of the kicked rotator,
the Floquet operator has matrix elements
\begin{eqnarray}\label{kickedU}
U_{l,l'} &=& M e^{-(\pi i/M) [(l+l_0)^{2}+(l'+l_0)^2]}
\\
& \times & \sum_m e^{2 \pi i m(l-l')/M}   
e^{-(iMK/2\pi) \cos(2\pi (m+m_0)/M)} \nonumber
\end{eqnarray}
with $l_0=p_0 M/2 \pi$ and $m_0=x_0 M/2 \pi$.

We restrict ourselves to the symmetric situation
with $N_{\rm R,L}=N$. A $2N \times 2N$ 
scattering matrix can be constructed
from the Floquet operator $U$ as \cite{fyodorov}
\begin{equation}\label{smatrix}
S(\varepsilon) = P [\exp(-i \varepsilon) - U (1-P^T P)]^{-1} U P^T,
\end{equation}
using a $2 N \times M$ projection matrix $P$ which
describes the coupling to the leads. 
Its matrix elements are given by
\begin{eqnarray}\label{lead}
 P_{n,m}=\left\{\begin{array}{ll}
1& \mbox{if $n=m \in \{m_i^{(R)} \} \bigcup \; \{m_i^{(L)} \}$},\\
0& \mbox{otherwise.}
\end{array}\right.
\end{eqnarray}
An ensemble of samples with the same microscopic parameters
can be defined by varying the position 
$\{m_i^{(R,L)} \}$, $i=1, \ldots, N$ 
of the contacts to the left and right leads for fixed 
$\tD/\tau_{\rm f}=M/2 N$ and $K$
[$\tau_{\rm f}$ is the time of flight through the system, in this particular
instance it is the time between kicks, see (\ref{krot})]. We calculate
the conductance from the scattering matrix, which we numerically
construct via an iterative procedure as in Refs.~\cite{TwoWL,Jac04}.

In the universal regime $\tEc/\tD=0$, Ref.~\cite{bardarson} found, for 
the $\beta=1$ to
$\beta=2$ crossover  a magnetoresistance given by
\begin{eqnarray}\label{rmt-g-b}
\delta g &=& \frac{1}{4} \;
\frac{1}{1+(m_0/m_c)^2} \;\; ; \;\; 
m_c = \frac{4 \pi}{\sqrt{M \tD} K},
\end{eqnarray}
which has to be compared to Eq.~(\ref{eq:gwl-lorentz}).
Our task here is to investigate the fate of (\ref{rmt-g-b}) as 
$\tEc/\tD$ increases, and in particular to check
our analytical predictions (\ref{eq:gwl-lorentz}) that
(i) the magnetoconductance is Lorentzian with
(ii) a width which is independent of $\tEc$, but
(iii) an amplitude which is suppressed
exponentially $\propto \exp[-\tEc/\tD]$.

\begin{figure}
\begin{center}
\hspace{-0.5cm}\includegraphics[width=7cm]{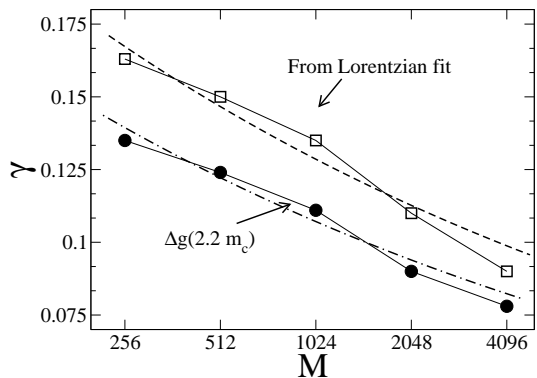}
\end{center}
\caption{\label{fignum2} 
Amplitude of the weak localization correction to the conductance
as a function of the size of Hilbert space. Data have been extracted 
from the curves shown on Fig.~\ref{fignum1} either via a 
fitting of the magnetoconductance curve with the Lorentzian
$\Delta g(m_0)=\gamma - \gamma/[1+\gamma' (m_0/m_c)^2]$ 
(squares) or taking $\gamma = \Delta g(2.2 m_c)$ (circles). Both methods 
confirm the exponential suppression $\gamma \propto \exp[-\tEc/\tD]$
(dashed and dotted-dashed lines) with an effectively smaller Lyapunov
exponent $\lambda \approx 1.26$ (compared to $\ln[K/2]\approx 1.95$) 
\cite{caveat3}.}
\end{figure}

Fig.~\ref{fignum1} shows magnetoconductance curves as $\tEc/\tD$ increases
while keeping all classical parameters unchanged. The data confirm our
prediction (\ref{eq:gwl-lorentz}), i.e. the curves are Lorentzian which depend
on $\tEc/\tD$ only through their amplitude. The inset of Fig.~\ref{fignum1}
makes it clear that the typical field necessary to break time-reversal 
symmetry does not depend on $\tEc$ --- after rescaling the
magnetoconductance amplitude, all curves fall on top of each other.
This was also confirmed by least square
Lorentzian fitting of the magnetoconductance curves 
with $\Delta g(m_0) = \gamma - \gamma/[1+\gamma' (m_0/m_c)^2]$,
which found $\gamma'=1 \pm 0.06$ for all cases.

Fig.~\ref{fignum2} finally gives a closer look at the suppression of the 
amplitude of the weak localization correction. We extracted the amplitude 
parameter $\gamma$ both from a least square Lorentzian fitting of the
magnetoconductance curve, 
and from the magnetoconductance amplitude at $m_0=2.2 m_c$, $g(2.2 m_c)-g(0)$.
The two procedures confirm the exponential suppression of
weak localization (\ref{eq:g_wl}), in agreement with Ref.~\cite{Bro05}.

\section{Conclusions}
\label{sect:conclusions}

In this paper we investigated the non-coherent transport properties
of open quantum chaotic systems in the semiclassical limit. We have
shown how to incorporate the nonergodic structures appearing in 
the classical phase-space (see Fig.~\ref{fig:cl_bands}) into 
quantum transport. We followed the scattering approach to transport
and showed how large phase-space structures result in a block-diagonal
form of the scattering matrix and the splitting of the system into
two sub-systems put in parallel. One of these sub-systems is of a purely
classical nature, consisting of deterministic transmission modes
(transmission eigenvalues are all 0 or 1). We were
able to calculate the corresponding transmission eigenvectors and connect the
emergence of determinism to the suppression of the Fano factor for shot-noise
as well as the breakdown of universality for sample-to-sample conductance
fluctuations.
The classical phase-space structures also cause the 
exponential suppression of 
coherent backscattering, preserving the unitarity of the
semiclassical theory of weak localization presented here.

At this point, the quantum mechanical subsystem is known not
to be RMT. 
Even though all its modes undergo a certain amount of mixing, and
thus carry quantum effects, weak localization and coherent 
backscattering come from trajectories longer than $\tEo+\tEc$, which have an
exponentially small relative weight $\exp[-\tEc/\tD]$ in the quantum cavity. 
The existence of two separated fluids is however confirmed. 

We finally point out that the phase-space method developed in the first half 
of this article
should work as well in regular systems. We however anticipate difficulties
not present in the 
chaotic systems treated here due to the power-law
decay of the band areas and diffraction effects at the leads. One 
open question is why open regular systems
with large dwell times have a RMT transmission spectrum
\cite{Marconcini04,Aigner05}, but a nonuniversal weak localization
behavior \cite{Bar93,Chang94}.

After this work was completed, a preprint appeared
which reached the same conclusions 
about coherent backscattering \cite{Rah06-cbs}.

\section*{Acknowledgments}

It is a pleasure to thank \.I. Adagideli and H. Schomerus for helpful
discussions, and C. Beenakker for drawing our attention to 
Ref.~\cite{bardarson}. This work has been financially supported by the 
Swiss National Science Foundation.
\appendix

\section{Gaussian wavepackets and hyperbolic dynamics}
\label{appendix:evol-gaussian-wavepackets}

We show that the time-evolution of a Gaussian wavepacket in a uniformly 
hyperbolic infinite system follows the Liouvillian flow.
This example supports the claim that 
a Gaussian wavepacket in a chaotic system
follows the Liouvillian flow up to the timescale at which
the wavepacket becomes so large 
that the Liouvillian dynamics ceases to be hyperbolic \cite{Heller}.

The uniformly hyperbolic Hamiltonian we consider is 
${\cal H} = p_x^2/(2m) - m\lambda^2x^2/2$. With the change of variable
$p_x= (m\lambda/2)^{1/2} (q+p_q)$ and $x = (2m\lambda)^{-1/2}(q-p_q)$, 
the Hamiltonian can be written as
$$
{\cal H} = \lambda \; (q p_q-\rmi \hbar/2) \ .
$$  
Solving the classical Hamilton equations of motion one gets
$q(t)=q(0)\e^{\lambda t}$ and $p_q(t)=p_q(0)\e^{-\lambda t}$.
Next, it is easily checked that 
a solution of the Schr\"odinger equation is provided by
the wavepacket
$$\langle q|\psi(t) \rangle 
= A \exp [-\Delta^{-2}(t)[q-q(t)-\rmi \Delta^2(t) 
p_q(t)/\hbar]^2],$$
with $\De(t)=\De (0)\e^{\lambda t}$. 
Thus we see that a Gaussian wavepacket remains Gaussian,
simply stretched and shifted by the Liouvillian flow.
This quantum calculation is exact if the system is infinite.
In finite systems, initially narrow 
classical distributions undergo a crossover
from hyperbolic dynamics to diffusive behavior once their extension 
become comparable to some characteristic length scale of the system \cite{ll}.
In our case we can thus expect that Gaussian wavepackets cease
to be Gaussian once the wavepacket has spread to a width of order 
the lead width.  In other words a Gaussian wavepacket will remain
Gaussian for times shorter than the Ehrenfest time \cite{Heller}.

\section{Algorithmic construction of an orthogonal 
phase-space basis}
\label{appendix:example-ps-basis}

\begin{table*}
\begin{tabular}{|r||c|c|c|c|c|c|c|c|c|c|c|} 
\hline
Iterations  & 
$\beta_0$ & $\beta_1$ & $\beta_2$ & 
$\beta_3$ & $\beta_4$ &  
$\beta_5$ & $\beta_6$ & $\beta_7$ & 
$\beta_8$ & $\beta_9$ & $\beta_{10}$\\
\hline\hline 
0 & 
1 & 0 & 0 & 0 & 0 & 0 & 0 & 0 & 0 & 0 & 0\\ 
\hline
2 & 
1.0386424 & -0.1137192 & 1.93720E-2 & -3.0781E-3 & 5.24E-6 & 
4.3E-6 & $\sim$0 & $\sim$0 &  $\sim$0 & $\sim$0 & $\sim$0 
\\ 
\hline
4 &  
1.0357044 & -0.1130793 & 1.74448E-2 & -3.0142E-3 &5.478E-4 &  
-1.024E-4 & 1.95E-5 & -3.8E-6 & 
7E-7 & -1E-7 & $\sim$0
\\
\hline
6 & 
1.0357044 & -0.1130793 & 1.74448E-2 & -3.0142E-3  & 5.478E-4 & 
 -1.024E-4 & 1.95E-5 & -3.8E-6 & 
7E-7 & -1E-7 & $\sim$0 
\\
\hline
10 &\  
1.0357044 \ &\  -0.1130793 \ &\ 1.74448E-2 \ &\  -3.0142E-3 \ &\ 
5.478E-4 \ &\  
-1.024E-4 \ &\  1.95E-5 \ &\  -3.8E-6 \ &\  
7E-7 \ &\  -1E-7 \ &\   $\sim$0 \   
\\ \hline
\end{tabular}
\caption{\label{table1}
Iterations of the orthogonalization algorithm for
coherent states on a von Neumann grid. 
The table shows the value of $\beta_i= \beta_{-i}$ 
as the algorithm is iterated ($\beta_j\sim 0$ means that 
$|\beta_j|< 10^{-7}$). The algorithm 
converges with accuracy $~10^{-7}$ after six iterations. Arbitrary large
accuracies are obtained with more iterations. The PS-states are found
by substituting the tabulated values into 
Eq.~(\ref{eq:example-ps-states}).}
\end{table*}

A basis of coherent states can be made complete (and not overcomplete)
but not orthogonal, by placing coherent states at the vertices of a
von Neumann lattice \cite{Perelomov}. This complete basis can be 
orthogonalized by following standard procedures 
(e.g. Gram-Schmidt orthogonalization), upon which,
however, the basis states become very different from one another
and extended in phase-space. We here describe a
numerical algorithm which orthogonalizes a complete basis of coherent
states on a von Neumann lattice, 
keeping all states identical (up to a translation in phase space)
while preserving the phase-space 
localization property of those states.

One starts from coherent states with wavefunctions
\begin{eqnarray}
|{\rm cs};i,j \rangle
= \exp \big[- \half |\alpha_{ij}|^2 \big] 
\exp \big[\alpha_{ij} \hat{a}^\dagger \big] |0\rangle . 
\label{eq:coh}
\end{eqnarray}
The creation and annihilation operators are (here $\hbar=1$)
\begin{eqnarray}
\hat{a}^\dagger = 2^{-1/2}[\hat{Q}-\rmi \hat{P}],
\qquad 
\hat{a} = 2^{-1/2}[\hat{Q}+\rmi \hat{P}],
\end{eqnarray}
where $\alpha_{ij} = 2^{-1/2}[Q_i+\rmi P_j]$ and
using dimensionless position $Q$ and momentum $P$ 
with $[\hat{P},\hat{Q}] = \rmi$.
The vacuum state, $|0\rangle$, is a Gaussian wavepacket centered
at $P=Q=0$,
\begin{eqnarray}
\langle Q | 0 \rangle = \langle Q | {\rm cs};0,0 \rangle 
&=& \pi^{-1/4} 
\exp [- \half Q^2 ].
\end{eqnarray}
The coherent state (\ref{eq:coh}) is a Gaussian 
wavepacket centered at $Q=Q_i$ and $P=P_j$
with the same spread in both directions,
\begin{eqnarray}
\langle Q | {\rm cs};i,j \rangle 
&=& \pi^{-1/4} 
\exp [\rmi P_j Q - \half[Q- Q_i]^2 ],
\nonumber\\
\langle P | {\rm cs};i,j \rangle 
&=& \pi^{-1/4} 
\exp [-\rmi P Q_i  - \half [P-P_i]^2 ], \qquad
\label{eq:Gaussians}
\end{eqnarray}
where we have dropped irrelevant overall phases.

To get a complete basis,
coherent states are placed at each vertex but one
of a square von Neumann lattice, i.e. a regular lattice on the $Q$-$P$
plane with each unit cell covering an area $(2 \pi)$ \cite{Perelomov}.
Translational invariance means that the empty lattice vertex
may be anywhere.
This basis of coherent states is complete but it is not 
orthogonal. To orthogonalize it, 
we make the ansatz that there exists a set $\{ \beta_i\}$
such that the wavefunction
\begin{eqnarray}
| {\rm ps};i,j \rangle = \sum_{i',j'}  \beta_{i'}\beta_{j'}
| {\rm cs};i'+i,j'+j \rangle
\label{eq:example-ps-states}
\end{eqnarray}
obeys 
\begin{eqnarray}
\langle{\rm ps};k,l | {\rm ps};i,j \rangle = \delta_{ik}\delta_{jl} \ .
\label{eq:condition-for-othonorm}
\end{eqnarray}
Note that the form of Eq.~(\ref{eq:example-ps-states}) is such that we 
assume that the basis-states will be symmetric under interchange of 
$Q$ and $P$ just as the coherent states are.

To satisfy Eq.~(\ref{eq:condition-for-othonorm}), we see that 
the elements of the set $\{ \beta_i\}$ must obey
\begin{eqnarray}
\delta_{i,0} \ = \ 
\sum_{i'i''} 
\beta_{i'}^*\beta_{i''} \ \exp [-(\pi/2)(i+i'-i'')^2]
\label{eq:condition-for-betas}
\end{eqnarray}
We define a set of vectors, $\{v^{(\al)} \}$, 
written in a non-orthogonal basis,
$\{ \hat{e}_i \}$, such that 
$v^{(\al)} = \sum_i \tilde{\beta}_{i-\al} \hat{e}_i$,
i.e. the $i$th element of the $\al$th basis vector is 
$v^{(\al)}_i = \beta_{i-\al}$.
The basis is chosen  such that the basis-vectors have the inner product  
$(\hat{e}_i \cdot \hat{e}_j) = \exp[-(\pi/2)(i-j)^2]$. 
The condition that the vectors  $\{v^{(\al)} \}$ form an orthonormal basis 
is 
\begin{eqnarray}
 \delta_{\al,0} 
&=& (v^{(\al)} \cdot v^{(0)})
= \sum_{ij} \beta_{i-\al}^* \beta_j  
(\hat{e}_i \cdot \hat{e}_j)
\end{eqnarray}
This is identical to the condition (\ref{eq:condition-for-betas}),
thus orthogonalizing this set of vectors is equivalent to
finding the $\beta$'s which satisfy Eq.~(\ref{eq:condition-for-betas}).
We use the following algorithm to orthogonalize these vectors
\begin{enumerate}
\item
Take a complete normalized (but non-orthogonal) basis, $\{ v_i \}$.

\item
define a new basis such  $\{ v'_i \}$ such that
$v'_i = A_i[v_i - \half \sum_{j\neq i} (v_i \cdot v_j) v_j]$.
We then choose $A_i$ such that it normalizes the vector $v'_i$.

\item 
Repeat the procedure, taking the new basis $\{ v'_i \}$
and deriving a basis $\{ v''_i \}$, and so on.
\end{enumerate}
We take the coherent states described above
as the initial non-orthogonal basis, so initially $\beta_i = \delta_{i,0}$.
In Table ~\ref{table1} we present data for the first ten
iterations of the algorithm, by the sixth iteration
the results satisfy Eq.~(\ref{eq:condition-for-betas})
with an accuracy of $\lesssim 10^{-7}$. 
Each iteration improves the accuracy by more than one order of magnitude.
A PS-state generated by this procedure is shown in Fig.~\ref{fig:ps-state}. 
We note that $\beta_i$ decays approximately exponentially with $i$.
Thus the PS-states given by Eq.~(\ref{eq:example-ps-states}) 
are exponentially localized in position and momentum,
as shown in Fig.~\ref{fig:ps-state}.

Area-preserving stretches,
$(Q,P,Q_j,P_j) \to (\kappa Q,\kappa^{-1} P,\kappa Q_j,\kappa^{-1} P_j)$,
and rotations  
$(Q,P,Q_j,P_j) \to 
(Q\cos\theta + P\sin\theta,P\cos\theta -Q\sin\theta,
Q_j\cos\theta + P_j\sin\theta, P_j\cos\theta -Q_j\sin\theta)$
are unitary operation for any $\kappa,\theta$.  
Thus the stretched-rotated basis will also be orthonormal and complete.
This legitimizes the procedure discussed in Section 
~\ref{sect:optimal-ps-basis}
for optimizing the PS-basis by fitting it to the PS scattering
band structure.

\section{Edge-of-band phase-space states}
\label{sect:appendix-eob-states}

The tails of the PS-state shown in Fig.~\ref{fig:ps-state}
decays exponentially with the number of lattice points away from
the center of the PS-state. 
Strictly speaking, any PS-state has thus a finite amplitude outside the
band. We can however treat PS-states
as classical, i.e. completely inside one band, if they
are more than $j_{\rm max}$ lattice sites away from the edge of that band.
If we choose $j_{\rm max}=1$, we would call states
``classical'' even if they have $\sim3\%$ of their 
squared amplitudes outside the band (this is similar to the situation
sketched in Fig.~\ref{fig:band-and-ps-states}). 
If however we take $j_{\rm max}=3$, then 
a PS-state is only classical if less than  $10^{-5}$ of its 
squared amplitude is outside the band.
The number of edge-of-band states 
(PS-states that are partially inside, partially outside a band
with area $>2\pi \hbar_{\rm eff}$) of a band which exits at time $\tau$ is 
\begin{eqnarray}
n^{{\rm L} \to K}_{\rm eob}(\tau) 
\simeq 4j_{\rm max}  
\left( { W_{\rm L} W_K \over 2\pi\hbar_{\rm eff}L^2} \right)^{1/2}
\exp [-\lambda \tau/2],
\end{eqnarray}
where $K={\rm L,R}$.  
Thus the number of edge-of-bands states is 
\begin{eqnarray}
N_{\rm eob} 
= \sum_K
\int_0^{\tau^{{\rm L}K}_{\rm E}} \! \rmd \tau
{\cal N}^{{\rm L}\to K}_{\rm band}(\tau)  n^{{\rm L}\to K}_{\rm eob}(\tau)
\, \sim \, {j_{\rm max}N_{\rm qm}\over \lambda\tau_{\rm D}}\qquad
\label{eq:N_eob-appendix}
\end{eqnarray}
where the sum is over $K={\rm L,R}$ and $j_{\rm max}$ is a number of order one.
Note that the error we make decays exponentially with $j_{\rm max}$, hence
the choice of acceptable error only changes $N_{\rm eob}$ logarithmically. 
From Eq.~(\ref{eq:N_eob-appendix})
we conclude that the edge-of-band PS-states are a subdominant proportion of 
the total number of quantum PS-states, and can be ignored.

\section{Flux enclosed by the correlated part of the loop}
\label{sect:magneto-correl}

\begin{figure}
\includegraphics[width=6cm]{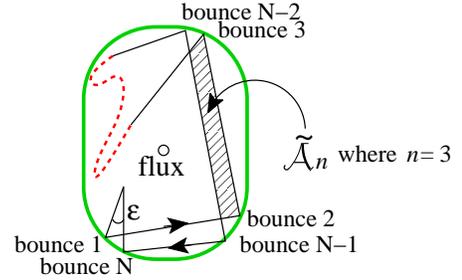}
\caption[]{\label{fig-magneto}
A sketch of the 
area enclosed by the correlated part of the weak localization loop.
The area $\tilde{\cal A}_n$ is defined by the
segment of the loop-forming trajectory between 
the $(n-1)$th and $n$th collisions and the segment between the
$(N-n+1)$th and the $(N-n+2)$th collisions. (dashed region).
}
\end{figure}

Here we analyze the part of the area enclosed by a loop-forming
 trajectory when it is in the correlation region close to the crossing,
i.e. within $T_L(\eps/2)$ of the crossing. The situation is
depicted in Fig.~\ref{fig-magneto}. We consider a loop formed 
after $N$ bounces at the cavity's wall.
In the correlation region, the segment of the  trajectory between 
the $(n-1)$th and $n$th collision with the cavity walls, 
is highly correlated with the segment between the $(N-n+1)$th
and the $(N-n+2)$th collision. We consider the directed area $\tilde{\cal A}_n$
enclosed by these two segments (dashed region in Fig.~\ref{fig-magneto}). 
We assume that $|\tilde{\cal A}_n|$ is uncorrelated
with  $|\tilde{\cal A}_{m\neq n}|$, and take each such area from a 
Gaussian distribution.
The typical perpendicular distance between the trajectories at time $t$ 
(measured from the crossing) is
$\pm v_{\rm F} \eps \lambda^{-1}\sinh (\lambda n\tau_0)$.
Thus we assume $\langle \tilde{\cal A}_n \rangle=0$ and
\begin{eqnarray}
\langle \tilde{\cal A}_n^2 \rangle
&=& \Big[ 
v_{\rm F}^2 \eps \lambda^{-1} 
\int_{(n-1)\tau_0}^{n\tau_0} \rmd t \sinh (\lambda t) \Big]^2.
\label{eq:typical-corr}
\end{eqnarray}
This grows exponentially with $n$.
Thus the sum over the $\tilde{\cal A}_n$'s is dominated by the largest
of them with $n_{\rm max} = T_L(\eps)/2\tau_{\rm f}$,
whose variance is $\sim A^2$. Anticorrelations in the signs
of consecutive directed areas in the correlated region further
reduce the total directed area. The flux enclosed in the
correlated region is thus at most $\simeq A^2 \Phi^2$. This is smaller than
the flux enclosed in the uncorrelated region by a factor 
$\tau_{\rm f}/\tD \ll 1$. 

\begin{thebibliography}{99}

\bibitem{ll} A.J. Lichtenberg and M.A. Liebermann, 
{\it Regular and Chaotic Dynamics},
(Springer, Berlin, 1983).

\bibitem{Eckmann} J.-P. Eckmann and D. Ruelle, Rev. Mod. Phys. {\bf 57},
617 (1985).

\bibitem{Wirtz99} L. Wirtz, J.-Z. Tang, and J. Burgd\"orfer,
Phys. Rev. B {\bf 59}, 2956 (1999); this 
paper discusses scattering bands in the context of quantum transport in
a chaotic billiard, but it does not make the connection between
their area and the quantum phase-space resolution $\hbar_{\rm eff}$.

\bibitem{Sil03} P.G. Silvestrov, M.C. Goorden, and C.W.J. Beenakker,
Phys. Rev. B {\bf{67}}, 241301(R) (2003).

\bibitem{TwoSN} J. Tworzydlo, A. Tajic, H. Schomerus and C.W.J. Beenakker, 
Phys. Rev. B {\bf 68}, 115313 (2003). 

\bibitem{TwoUCF} J. Tworzydlo, A. Tajic and C.W.J. Beenakker, Phys. Rev. B
  {\bf 69}, 165318 (2004).

\bibitem{Jac04} Ph. Jacquod, and E.V. Sukhorukov, Phys. Rev Lett. {\bf{92}}
116801 (2004). 

\bibitem{Goo05} M.C. Goorden, Ph. Jacquod, and C.W.J. Beenakker,
Phys. Rev. B {\bf 72}, 064526 (2005).

\bibitem{caveat1} 
The dwell time is exactly the same for all trajectories
within a band for maps such as the kicked rotator used in 
Figs.~\ref{fig:cl_bands} and \ref{fig:qm_bands}.

\bibitem{Revdot} L.P. Kouwenhoven, C.M. Marcus, P.L. McEuen, S. Tarucha, R.M. 
Westervelt, and N.S. Wingreen,
{\it Electron Transport in Quantum Dots},
Nato ASI conference proceedings, 
L.P. Kouwenhoven, G. Sch\"on, and L.L. Sohn Eds. 
(Kluwer, Dordrecht, 1997);
Y. Alhassid, Rev. Mod. Phys. {\bf 72}, 895 (2000);
I.L. Aleiner, P.W. Brouwer, and L.I. Glazman, Phys. Rep. {\bf 358},
309 (2002).

\bibitem{Bar93} H.U. Baranger, R.A. Jalabert, and A.D.
Stone, Chaos {\bf 3}, 665 (1993).

\bibitem{Mar93} C.M. Marcus, R.M. Westervelt, P.F. Hopkings, and 
A.C. Gossard, Chaos {\bf 3}, 643 (1993). 

\bibitem{Meh91} M. L. Mehta, {\em Random Matrices} (acadamic, New York, 1991).

\bibitem{RMT} R.A. Jalabert, J.-L. Pichard, and C.W.J. Beenakker,
Europhys. Lett. {\bf 27}, 255 (1994);
H.U. Baranger and P.A. Mello, Phys. Rev. Lett. {\bf 73},
142 (1994). 

\bibitem{haake} S. Heusler, S. M\"uller, 
P. Braun, and F. Haake, Phys. Rev. Lett. {\bf 96}, 066804 (2006) .

\bibitem{Berman78} G.P. Berman and G.M. Zaslavsky, Physica {A \bf 91},
450 (1978).

\bibitem{Obe02} S. Oberholzer, E.V. Sukhorukov, and C.
Sch{\"o}nenberger, Nature (London) \textbf{415}, 765 (2002).

\bibitem{Bla00} Ya.M. Blanter and M. B\"uttiker, Phys. Rep. {\bf 336}, 1 
(2000).

\bibitem{vanhouten} C.W.J. Beenakker 
and H. van Houten, Phys. Rev. B {\bf 43}, R12066 (1991); the 
semiclassical limit requires that one adapts the voltage
in order to fix the injection rate of electrons, so
$V \propto N_{\rm L}^{-1}$ as $\hbar_{\rm eff} \to 0$. 
If the voltage were fixed, the shot-noise power would diverge as
$\hbar_{\rm eff} \to 0$.
We thank Carlo Beenakker for a discussion of this point.

\bibitem{Aga00} O. Agam, I. Aleiner and A. Larkin, Phys. Rev. Lett.
{\bf 85}, 3153 (2000).

\bibitem{wj-fano} R.S. Whitney and Ph. Jacquod, cond-mat/0512516.

\bibitem{Whi05} R.S. Whitney, and Ph. Jacquod, Phys. Rev. Lett. {\bf 94},
116801 (2005).

\bibitem{Henning05} For a recent review on the
quantum-to-classical correspondence in open systems see:
H. Schomerus and Ph. Jacquod, J. Phys. A. {\bf 38}, 10663 (2005).

\bibitem{Ale96} I.L. Aleiner and A.I. Larkin, Phys.\ Rev.\ B {\bf 54},
14423 (1996).

\bibitem{Ada03} \.I. Adagideli, Phys.\ Rev.\ B {\bf 68}, 233308 (2003).

\bibitem{TwoWL} J. Tworzydlo, A. Tajic and C.W.J. Beenakker, Phys. Rev. B 
{\bf 70}, 205324 (2004).

\bibitem{Bro05} S. Rahav and P.W. Brouwer, Phys. Rev. Lett. {\bf 95}, 056806 
(2005); cond-mat/0507035.


\bibitem{Vav03} M.G. Vavilov and A.I. Larkin, Phys.\ Rev.\ B {\bf 67},
115335 (2003).

\bibitem{Bauer90} W. Bauer and G.F. Bertsch, Phys. Rev. Lett. {\bf 65},
2213 (1990).

\bibitem{Scho04} H. Schomerus and J. Tworzydlo, Phys. Rev. Lett. {\bf 93},
154102 (2004).

\bibitem{wavelet-books}
I. Daubechie, {\em Ten Lectures on Wavelets} (SIAM, Philadelphia, 1992);
G. Kaiser, {\em A Friendly Guide to Wavelets} (Birkh\"auser, Boston, 1994).

\bibitem{footnote:square-ps-basis}
If the L lead is covered with PS-states placed on a square grid,
then each PS-state has a spread 
of $\hbar_{\rm eff}^{1/2}$ in $r$ and $p$.
Thus a classical band which is narrower than 
$\hbar_{\rm eff}^{1/2}$ could not contain any PS-states,
despite the band's phase-space area being
$\sim \hbar_{\rm eff}^{1/2}W/L \gg \hbar_{\rm eff}$.
This would lead to the prediction that the open cavity Ehrenfest time
is $\lambda^{-1}\ln[\hbar_{\rm eff}^{1/2}W/L]$
(for $W_{\rm L}=W_{\rm R}=W$), which
is half the value found by using the optimized PS-basis, see 
Eq.~(\ref{eq:tau_E^LK}).
\bibitem{footnote:lambda}
In Eq.~(\ref{eq:aspect-ratio}), 
$\lambda$ is, strictly speaking, not the Lyapunov exponent, 
but rather the local rate of stretching of the Liouvillian flow
averaged along the given trajectory. This distinction is important for
the covering of non-parallelogram bands.

\bibitem{Heller} E.J. Heller and S. Tomsovic, Phys. Today {\bf 46}(7),
38 (1993).

\bibitem{fisher_lee} D.S. Fisher and P.A. Lee, Phys. Rev. B {\bf 23},
R6851 (1981).

\bibitem{footnote:delta_hbar}
For an ideal lead, with $N$ lead modes of the form
$\langle y|n\rangle= (2/W)^{1/2} \sin (\pi yn/W)$ for $0\leq y \leq W$,
one finds that 
$\sum_n\langle y'|n\rangle\langle n|y\rangle
=(2W)^{-1}[ \sin [(z'-z)(N+1/2)] /\sin [(z'-z)/2]
- \sin [(z'+z)(N+1/2)] /\sin [(z'+z)/2]]$
 where $z=\pi y/W$.
This function is strongly peaked at $y'=y$
with peak width $\sim\lambda_{\rm F}$ and height $\sim\lambda_{\rm F}^{-1}$,
in the semiclassical limit we can calculate all quantities
to lowest order in $\hbar$ by approximating this function with a Dirac 
$\de$-function.

\bibitem{Ric02} K. Richter and M. Sieber, Phys. Rev. Lett. {\bf 89}, 
206801 (2002).

\bibitem{Ric01} M. Sieber and K. Richter, Phys. Scr. {\bf T90}, 128 (2001).

\bibitem{caveat2} Unlike in Section~\ref{sect:Ehren-times}, 
we neglect the lead asymmetry 
in the definition of $T_W(\epsilon)$, which would lead to corrections
of order $\sim \ln|\WL/\WR|$ only. Neglecting them is consistent with
our treatment of weak localization which
neglects other corrections of order one inside logarithms of the
Ehrenfest and $\epsilon-$times. See the remark before Eq.~(\ref{eq:cbs}).

\bibitem{footnote:mea-culpa}
This is the origin of our incorrect previous
prediction that $g_{\rm wl}$ would not decay with $\tEc$ \cite{Whi05}.
Our reasoning was that the probability to return should be 
the inverse volume of the quantum phase-space rather than the total 
phase-space.  That reasoning is incorrect, as shown by the present
calculation where crossings contributions are clearly identified.
Here, one is left with probabilities for classical trajectories 
which know nothing of $\hbar$ or $\tEc$.  
The suppression we expected due to the existence of
bands is fully accounted for in the correlated escape probabilities for 
almost parallel  trajectories  within $W$ of each other,
it reduces the decay from $\exp[-(\tEc+\tEo)/\tD]$ to  $\exp[-\tEc/\tD]$.


\bibitem{izrailev} F. M. Izrailev, Phys.\ Rep.\ {\bf 196}, 299 (1990).

\bibitem{bardarson} J.H. Bardarson, J. Tworzydlo, and C.W.J. Beenakker, 
Phys. Rev. B {\bf 72}, 235305 (2005).

\bibitem{fyodorov} Y.V. Fyodorov and H.-J. Sommers,
JETP Letters {\bf 72}, 422 (2000); R.O. Vallejos and A.M.
Ozorio de Almeida, Ann. Phys. \textbf{278}, 86 (1999).

\bibitem{caveat3} Our semiclassics implicitly averages over
encounters distributed all over phase-space. The exponential suppression
of weak localization is thus $\propto \langle \exp[-\tEc/\tD] \rangle$.
For systems which are nonuniformly hyperbolic such as the kicked rotator, 
the ``Lyapunov exponent'' measured by the decay of weak-localization 
might thus differ from $\ln[K/2] = \langle \lambda \rangle$. 
A similar effect has been found in the decay of the fidelity in
dynamical systems, P.G. Silvestrov, J. Tworzydlo, and C.W.J. Beenakker,
Phys. Rev. E {\bf 67}, 025204(R) (2003); C. Petitjean and Ph. Jacquod,
Phys. Rev. E {\bf 71}, 036223 (2005).

\bibitem{Marconcini04}
P. Marconcini, M. Macucci, G.
Iannaccone, B. Pellegrini, and G. Marola, 
Europhys. Lett. {\bf 73}, 574 (2006).

\bibitem{Aigner05} F. Aigner, S. Rotter, and J. Burgd\"orfer, Phys.
Rev. Lett. {\bf 94}, 216801 (2005).

\bibitem{Chang94}
A.M. Chang, H.U. Baranger, L.N. Pfeiffer, and K.W. West,
Phys. Rev. Lett. {\bf 73}, 2111 (1994).

\bibitem{Rah06-cbs}
S.~Rahav and P.W.~Brouwer, Phys.~Rev.~Lett.~{\bf 96}, 196804 (2006).


\bibitem{Perelomov} A. Perelomov, {\it Generalized 
Coherent States and Their Application}, 
(Springer-Verlag, Berlin, 1986). 

\end{thebibliography}
\end{document}